\documentclass[useAMS,usenatbib,usegraphicx]{mn2e}
\usepackage{epsf,rotating,url}
\usepackage{subfigure,amssymb,amsfonts,amstext,amsgen,amsopn,amsxtra,indentfirst,lscape,times,rotating}
\usepackage{longtable}
\usepackage{graphics}
\usepackage{keyval}
\usepackage{trig}
\usepackage{dcolumn}
\usepackage{bm}
\def\beq{\begin{equation}}
\def\eeq{\end{equation}}
\def\bey{\begin{eqnarray}}
\def\eey{\end{eqnarray}}
\def\msun{M_\odot}

\def\kms{\, {\rm km \, s}^{-1} }

\def\mnras{MNRAS}
\def\apj{ApJ}
\def\nat{Nature}
\def\apjs{ApJ}
\def\apjl{ApJ}
\def\na{New Astronomy}
\def\araa{ARAA}

\def\aap{A \& A}
\def\aj{AJ}
\def\lcdm{{\Lambda}CDM}

\def\aap{Astron. Astrophys.}
\def\jcap{JCAP}

\title{Using dwarf satellite proper motions to determine their origin}

\author[G. W. Angus, A. Diaferio, P. Kroupa]{G. W. Angus$^{1,2,3}$\thanks{E-mail: angus.gz@gmail.com}, Antonaldo Diaferio$^{2,3,4}$ and Pavel Kroupa$^5$ \\ 
$^{1}$Astrophysics, Cosmology \& Gravity Centre, University of Cape Town, Private Bag X3, Rondebosch, 7700, South Africa \\
$^{2}$Dipartimento di Fisica Generale ``Amedeo Avogadro", Universit\`a degli studi di Torino, Via P. Giuria 1, I-10125, Torino, Italy \\
$^{3}$Istituto Nazionale di Fisica Nucleare (INFN), Sezione di Torino, Torino, Italy\\
$^{4}$Harvard-Smithsonian Center for Astrophysics, 60 Garden Street, Cambridge, MA 02138, USA\\
$^{5}$Argelander Institute for Astronomy, University of Bonn, Auf dem H\"ugel 71,D-53121 Bonn, Germany}
\begin{document}

\date{\today}
\maketitle
\begin{abstract}
The highly organised distribution of satellite galaxies surrounding the Milky Way is a serious challenge to the concordance cosmological model. Perhaps the only remaining solution, in this framework, is that the dwarf satellite galaxies fall into the Milky Way's potential along one or two filaments, which may or may not plausibly reproduce the observed distribution. Here we test this scenario by making use of the proper motions of the Fornax, Sculptor, Ursa Minor and Carina dwarf spheroidals, and trace their orbits back through several variations of the Milky Way's potential and account for dynamical friction. The key parameters are the proper motions and total masses of the dwarf galaxies. Using a simple model we find no tenable set of parameters that can allow Fornax to be consistent with filamentary infall, mainly because the $1\sigma$ error on its proper motion is relatively small. The other three must walk a tightrope between requiring a small pericentre (less than $20~kpc$) to lose enough orbital energy to dynamical friction and avoiding being tidally disrupted. We then employed a more realistic model with host halo mass accretion, and found that the four dwarf galaxies must have fallen in at least 5~Gyrs ago. This time interval is longer than organised distribution is expected to last before being erased by the randomisation of the satellite orbits.
\end{abstract}

\begin{keywords}
galaxies: dwarf – galaxies: evolution – gravitation – Local Group – dark matter
\end{keywords}
\section{Introduction}
\protect\label{sec:intr}
The observed number, spatial distribution and orbital energy of satellite galaxies surrounding the Milky Way is quite surprising in terms of the concordance cosmological model (see e.g. \citealt{spergel07}), where galaxy hosting cold dark matter halos are formed by the hierarchical merging of many subhalos.  Before any baryonic physics has been considered, we would expect a moderately anisotropic distribution of cold dark matter subhalos surrounding the main halo, like shown in \cite{lihelmi08}, that number in the hundreds (see in particular \citealt{klypin99,moore99}). This prediction is clearly at odds with the small $\sim$ O(10) number of observed satellites - which should not include the unbound Magellanic Clouds: see e.g. \cite{kallivayalil06b}, \cite{kallivayalil06a} and \cite{piatek08}.

The perfectly mundane solution to this problem is that the majority of the satellites are not efficient at forming stars and so stay dark. This is more or less achievable if one tunes the available free parameters surrounding the feedback of supernovae and other poorly understood astrophysical processes (\citealt{efstathiou00,madau01,springel03,veilleux05,governato07} amongst many others) contributing to star and galaxy formation.

This is not the end of the story, however, because the spatial distribution of these remaining satellites is equally puzzling. As highlighted in \cite{kroupa05,metzkroupa07} and previously noted by \cite{lb83}, their current positions around the Milky Way can be likened to an extended disk with a root mean squared thickness of merely $10-30~kpc$. This is again contrary to the naive expectation in the concordance cosmological model because {\it a priori} one would expect the luminous satellites to be randomly distributed.

This argument has been countered by \cite{lihelmi08} (see also \citealt{dong08}) who claimed the thin nature of the disk of satellites (DoS) might be explained by the infall of a small group of galaxies that would retain the correlated orbits. This suggestion was dissected by both \cite{klimentowski10} and \cite{metz09} who both gave arguments as to why the model would likely fail. In particular, \cite{metz09} pointed out that no nearby groups are observed to be anywhere near as flattened as the disk of satellites, and \cite{klimentowski10} showed, with N-body simulations, that the coherence timescale for these groups is short lived after infall.

It is worth keeping in mind, however, that the Milky Way is not isolated and that the presence of M31 in our vicinity may have some bearing on the formation of the Milky Way system as a whole (see e.g. \citealt{sawa05,pasetto07}). Also there is the possibility of an intricate set of circumstances whereby several dwarfs in a dynamical group are sacrificed to allow the survival of single member through scattering (\citealt{sales07a,sales07b}). However, perhaps besides these, the only remaining solution is that of \cite{kang05,libeskind05,libeskind07,libeskind09,zentner05} who have suggested that the satellites fall in at high redshift along filaments which creates a preferred orbital direction and might satisfy the small width of the DoS. However, there is a further piece of the jigsaw that has been neglected, or overlooked until now, and that is the proper motions of the dwarf satellites. When these dwarfs fall along the filament, they accelerate towards the Milky Way and, to match the observed proper motions, they must lose orbital energy along the way, presumably to dynamical friction.

This is important because the proper motions of four of the classical dwarf spheroidal galaxies (dSphs) - Fornax, Sculptor, Ursa Minor and Carina - have been measured by \cite{piatek07,piatek06,piatek05,metz08} respectively. In order for the filamentary infall hypothesis to be valid it must be possible for satellites to fall in to some realistic Milky Way potential and at some point reach the current positions of each of the dSphs and be consistent with their measured proper motions and radial velocities. The purpose of this paper is to test this argument.

\section{Initial conditions and strategy}

The strategy we employ to test the filamentary infall scenario is quite simple. 

\subsection{Dynamical Friction}

We modified an orbit integrator, which usually simply updates the position and velocity of a test particle according to some gravitational potential, to include the effects of dynamical friction. Dynamical friction (DF) is the retarding force experienced by a massive object as it orbits through a sea of particles or stars. As it moves, the object scatters oncoming particles creating an overdense wake to its rear. This overdensity acts gravitationally to decelerate the massive object. According to \cite{bt08}, the deceleration due to DF can be parametrised as follows

\bey
\nonumber a_{df}={-2\pi \ln(1+ \Lambda^2) G^2\rho_{dm}(r) M_d \over V^2}\times\\
\left[erf (V/v_c(r))-{2 \over \sqrt{\pi}}{V \over v_c(r)}e^{-V^2/v_c(r)^2}\right].
\protect\label{eqn:df}
\eey
Here, $M_d$ is the total mass of the dSph and $V$ is the velocity of the dSph with respect to the Milky Way; $G$ is Newton's constant, $v_c(r)$ is the local circular speed defined by the Milky Way potential, $\rho_{dm}(r)$ is the local dark matter density (we assume that the dSphs orbit outside the disk) and $\ln (1+\Lambda^2)$ is the Coulomb Logarithm. Ordinarily, $\ln (1+\Lambda^2)$ can be reduced to $2\ln \Lambda$, but in our case $\Lambda$ can be quite small, so it is better left unaltered.

In addition to this, $\Lambda$ is often parametrised as ${b_{max}V \over G M_d}$ where $b_{max}$ is linked to the maximum impact parameter of particles that meaningfully contribute to the deceleration. In this work we take $b_{max}=20~kpc$, which is $20-40$ times the current, stellar tidal radii of the four dSphs. To gauge how important this parameter is, we note that for $M_d=10^9$, $10^{10}$ and $10^{11}\msun$, with $V=300\kms$ we can double $b_{max}$ and only induce a 10\%, 20\% and 45\% respective increase in the Coulomb logarithm. Therefore, if one wishes to consider a larger $b_{max}$ than we do here, simply surmise that the true $M_d$ is slightly lower than the used one. However, for realistic dSph masses ($<10^9\msun$) this effect is small.
\subsection{Milky Way mass model}
\protect\label{sec:4m}
The dSphs mainly orbit in the halo of the Milky Way at Galactocentric radii $> \sim 20~kpc$, therefore, the inner rotation curve - which is a hotly debated topic (see e.g. \citealt{bovy09} and references therein) - is of little concern. The main constraint we are interested in is the rotation curve measured at large radii from the Milky Way. The two articles we pay particular attention to are those of \cite{gnedin10} and \cite{xue08}.

We use the same potentials as \cite{xue08} for the bulge, disk and dark matter halo and we do not repeat them here for brevity. We choose $M_{disk}=5\times10^{10}\msun$, $M_{bulge}=1.5\times10^{10}\msun$, but like \cite{mcgaugh08} we choose the scale-length of the disk to be smaller (in this case $b=2.8~kpc$, not $4~kpc$ as \citealt{xue08} choose) in order to increase the rotation speed near the solar radius to the expected $v_c(r_{\sun})\approx 220~\kms$ (see \citealt{mcgaugh08}). Of course we do not need it to be particularly accurate since we are using a spherical potential to model the disk and the fact is it does not influence the orbits of dSphs at $20~kpc$, never mind beyond $100~kpc$.

Given that we are testing the $\lcdm$ model, it makes no sense to use a logarithmic halo, so we only use a \citealt{nfw97} halo (hereafter NFW). We note, however, that a logarithmic halo has by definition stronger gravity in the outer parts of the halo and so the infalling satellites accelerate to larger speeds than with an NFW halo, which is counterproductive to the need to lose orbital energy, as we shall see later.

\begin{figure*}
\centering
\subfigure{
\includegraphics[angle=0,width=5.50cm]{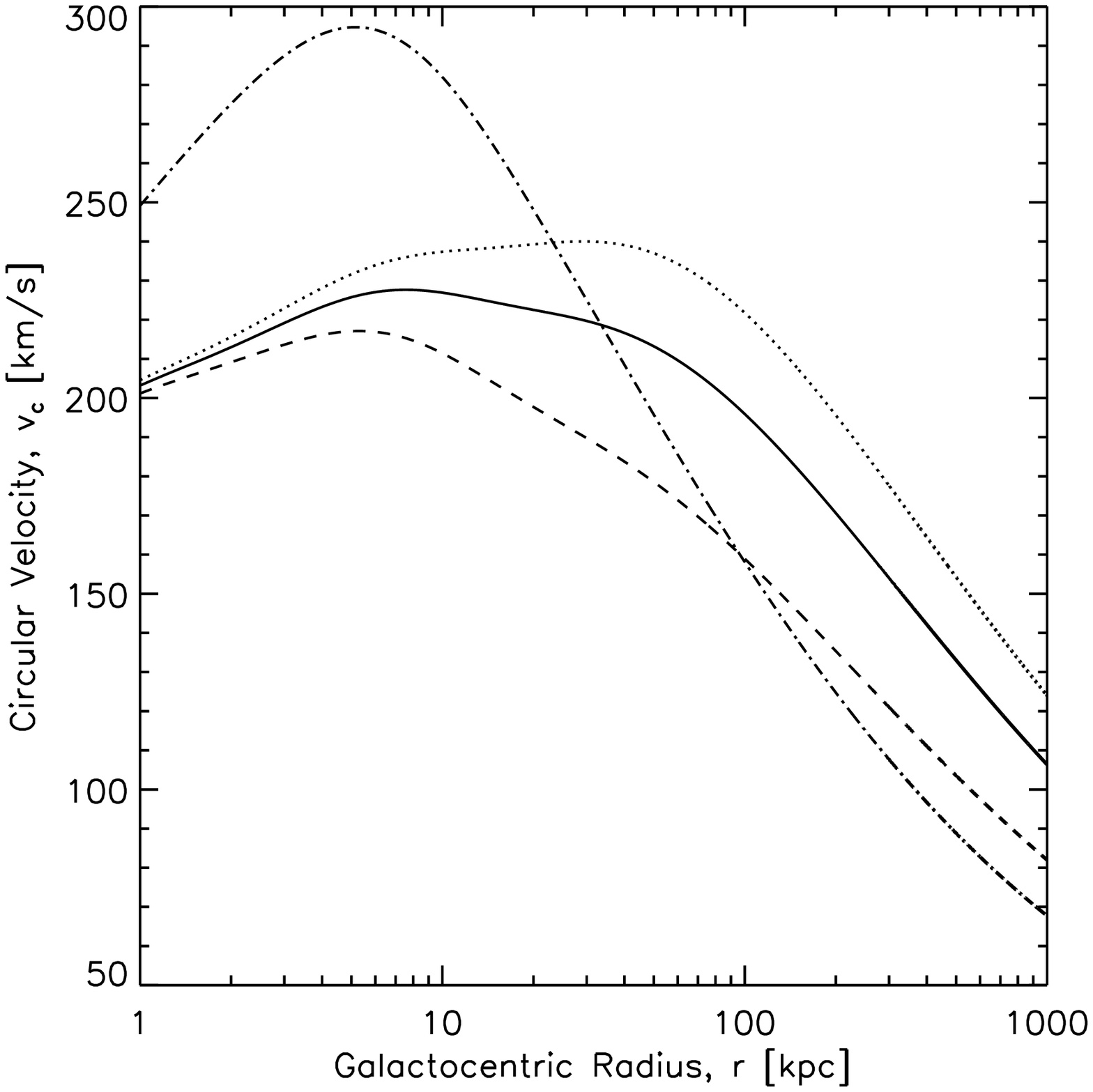}
}
\subfigure{
\includegraphics[angle=0,width=5.50cm]{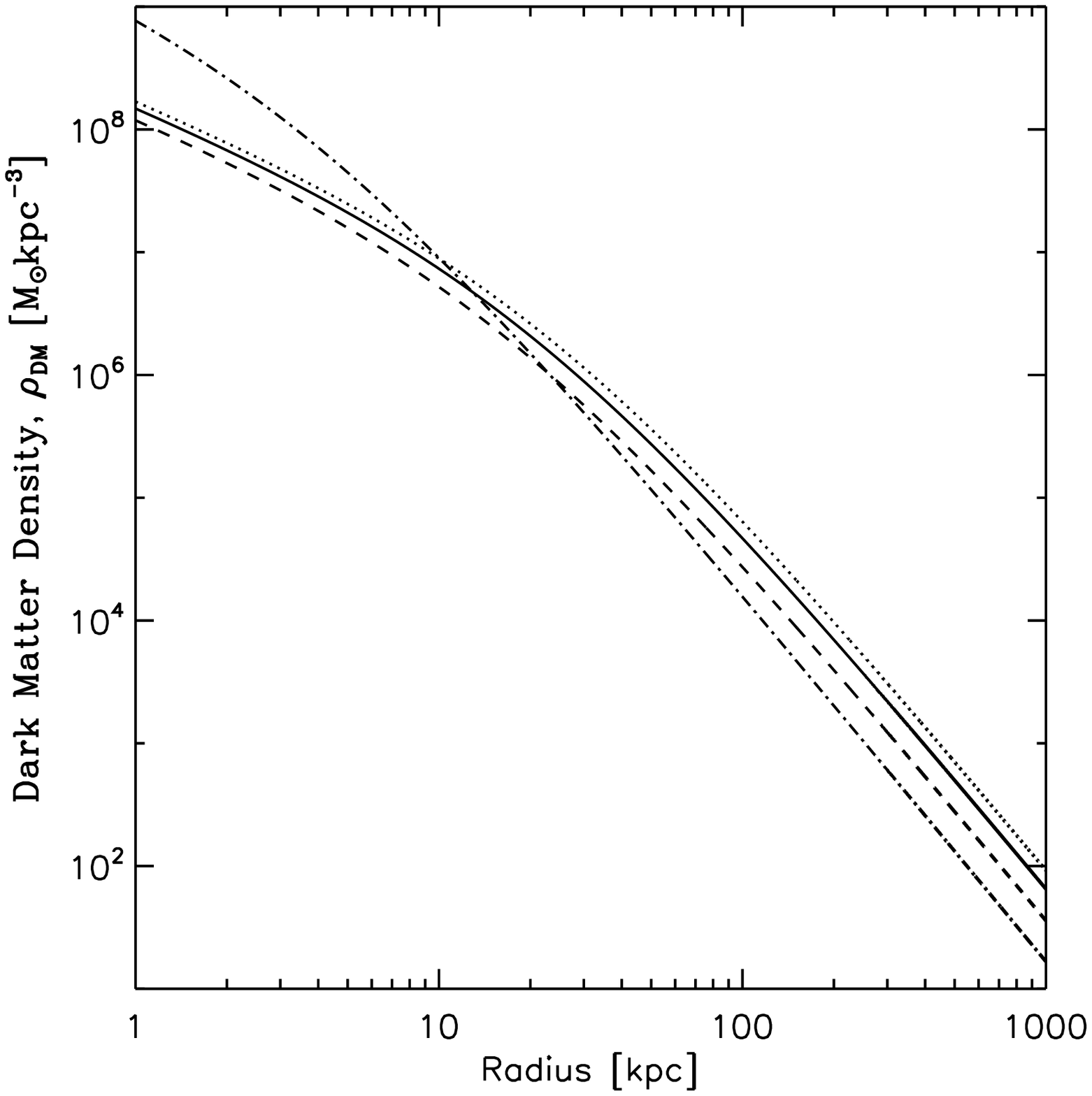}
}
\subfigure{
\includegraphics[angle=0,width=5.50cm]{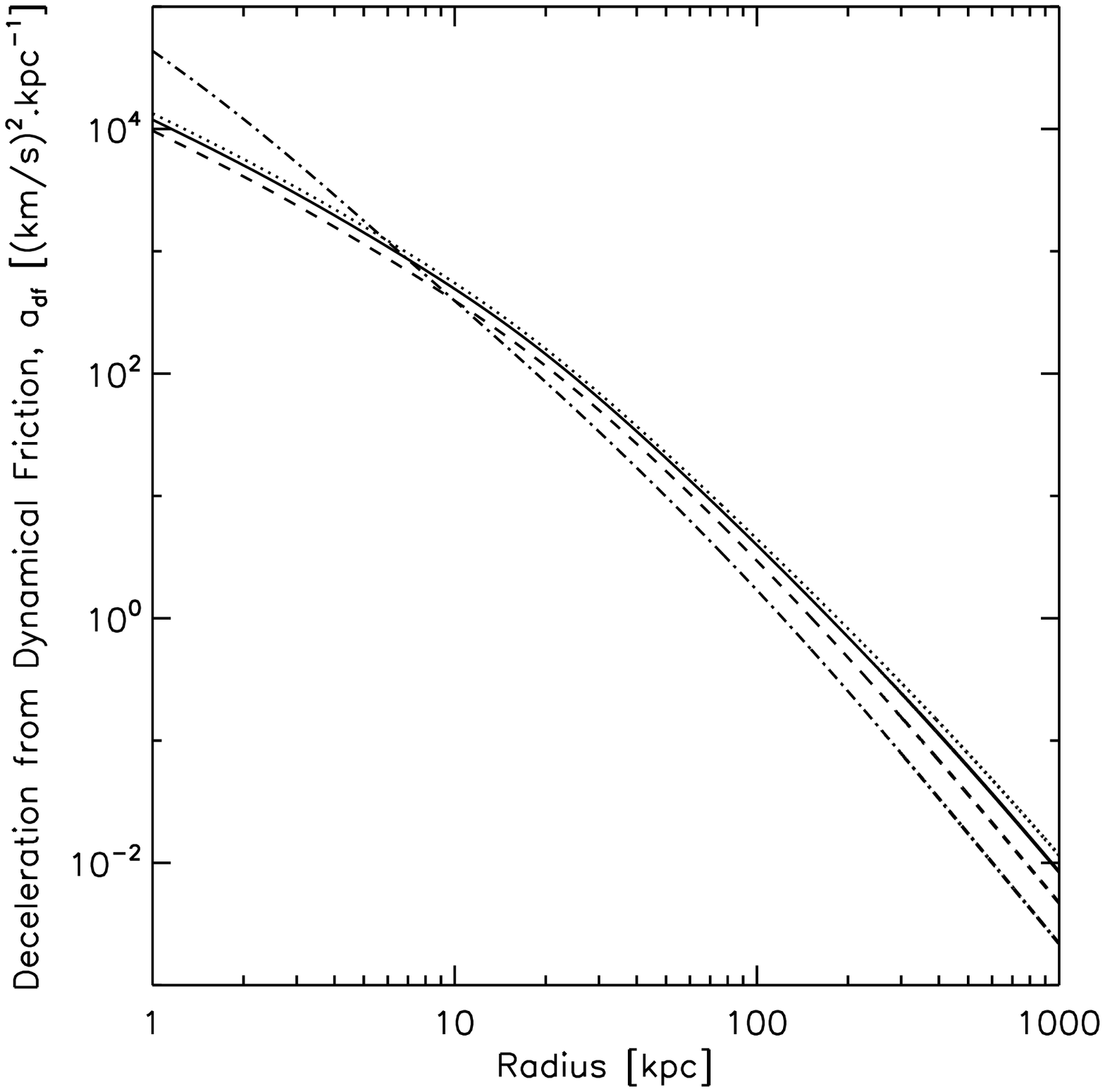}
}\\
\caption{A series of important quantities are plotted against Galactocentric radius for the four different Milky Way mass models ( where the only differences are the dark matter halos). In panel (a) is the circular velocity out to 1~$Mpc$, (b) shows the dark matter density and panel (c) shows the deceleration due to dynamical friction for a dwarf satellite mass of $10^{10}\msun$ moving at 250~$\kms$. The linestyles for models I, II, III, IV are solid, dotted, dashed and dot-dashed respectively.}
\label{fig:vc}
\end{figure*}

To give the extremes of the permissible rotation curves, we take Fig~3 of \cite{gnedin10} which plots the allowed range of rotation curves between 20 and 80~$kpc$. From this we chose four halo models: one to represent a rotation curve that is roughly in the centre of the allowed range of velocities from 20-80~$kpc$ (model I); one to represent the largest viable rotation curve (model II), another to represent the lowest viable rotation curve (model III) and a final one that falls from a large rotation speed at 20~$kpc$ to a low rotation speed at 80~$kpc$ (model IV). The NFW concentration parameters, $c$, and densities, $\rho_s$, for the four models are given in Table~1 and we note that the virial radius at redshift zero, $r_{200}$, is fixed to 275~$kpc$. Here $r_{200}$ refers to the radius at which the enclosed mass is two hundred times the critical density of the Universe, $M_{200}={200 \rho_c \times {4 \over 3}\pi r_{200}^3}$. We vary only the concentration parameter, $c$, for models I-III, but also the density, $\rho_s$ for model IV, where $\rho=\rho_s{(r_{200}/c)^3 \over {r(r_{200}/c+r)^2}}$. The rotation curves of the four models can be seen in Fig~\ref{fig:vc} as well as their dark matter densities and their typical DF decelerations for some reasonable parameters.

Finally, increasing-mass Milky Way models are introduced in \S\ref{sec:hha} to account for the infall of dark matter onto the early Milky Way.

\subsection{dSph Proper Motions}
The crux of the whole issue revolves around the knowledge of the proper motions of the four dSphs, and of course their heliocentric radial velocities and distances from the Galactic centre  (see e.g. \citealt{mateo98}). The Galactocentric distances to the four dSphs are given in Table~\ref{tab:pm} and all have relatively small errors, so we do not admit any variation in these current positions.

The main uncertainty in accounting for the past orbits of the dSphs comes from the proper motions. The proper motions of Fornax (\citealt{piatek07}), Sculptor (\citealt{piatek06}), Ursa Minor (\citealt{piatek05}) and Carina (\citealt{piatek03}), measured in those papers, were converted into Galactic rest frame tangential and radial velocities (by the original authors) and are given in Table~\ref{tab:pm}. Clearly there is a significant amount of uncertainty associated with these velocities, and they have the potential to drastically alter the orbits.

\subsection{Orbit Evolution}
Our strategy, given the information available, is to elucidate the likelihood that the four dSphs fell into the Milky Way along a filament and ended up with their current positions and velocities. To do this, we begin from their current positions and velocities and trace back their orbits (accounting for DF) to see if, for a given dSph mass and Galactic potential, it is possible for the dSph to reach a significant distance from, or even escape, the Milky Way.


The equations of motion are given in the appendix.

\section{Results}
Our condition for a viable dSph candidate arriving by filamentary infall is the following: when we trace its orbit back in time, starting from its current position and range of possible velocities, it must be able to reach a significant distance from the Milky Way within $12~Gyr$. The maximum trace back time is curtailed to $12~Gyr$ because any dSphs falling in before $12~Gyrs$ would have their positions with respect to the other dSphs randomised such that they could not account for the disk of satellites. This distance must be substantially greater than the current orbital apocentre found assuming a fixed orbit with no changing Galactic potential and no dynamical friction. If this were satisfied, then it would suggest that dynamical friction is strong enough to decelerate an infalling the dSph and allow it to be consistent with its current proper motions.

The variables discussed above include the Galactic potential, the proper motions and dSph mass. We have four Galactic potentials (parameters in Table~\ref{tab:mod}), vary the proper motions between $-3\sigma$ and $+3\sigma$ of their mean measurements and consider dSph masses up to $10^{12}\msun$. The aim is to find the minimum mass permissible, from constraints on the Galactic potential and proper motions, and then to judge the credibility of such a mass given the information we have on dSph masses from recent, independent observations.

\begin{figure}
\centering
\subfigure{
\includegraphics[angle=0,width=8.0cm]{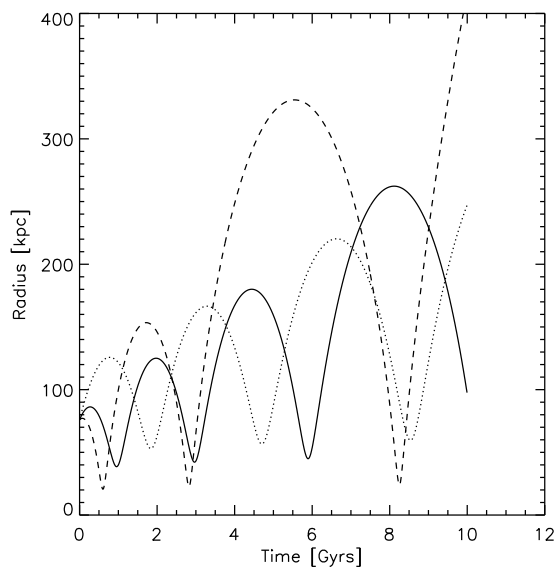}
}\\
\caption{Here we show sample orbits for the Ursa Minor dwarf spheroidal galaxy (using a $5\times 10^{10}\msun$ mass and Galactic model I) to demonstrate how the orbital apocentre decays with time due to dynamical friction. The different lines are for different values of the observed proper motion. The solid line is the mean, the dotted line is found by adding the $1\sigma$ error in parallel to the mean and the dashed line is adding the $1\sigma$ error anti-parallel.}
\label{fig:orbit}
\end{figure}

In Fig~\ref{fig:orbit} we show the decay of a set of orbits for the Ursa Minor dSph for various proper motions and using Galactic model I. It shows how critical the pericentre is for allowing the  dSph to lose enough orbital energy to satisfy the current position and velocity after infall. Although the solid and dotted lines (mean measurement and mean+$1\sigma$ error in proper motions) do not efficiently lose their orbital energy, the dashed line (mean subtracting the $1\sigma$ error) does. Therefore, if Ursa Minor was observed to weigh upwards of $5 \times 10^{10}\msun$ and if the proper motions were confirmed to be $1\sigma$ lower than the current mean measurement, then it would be a good candidate for filamentary infall.

\subsection{Dependence on the Galactic Model}
\begin{figure*}
\centering
\subfigure{
\includegraphics[angle=0,width=8.0cm]{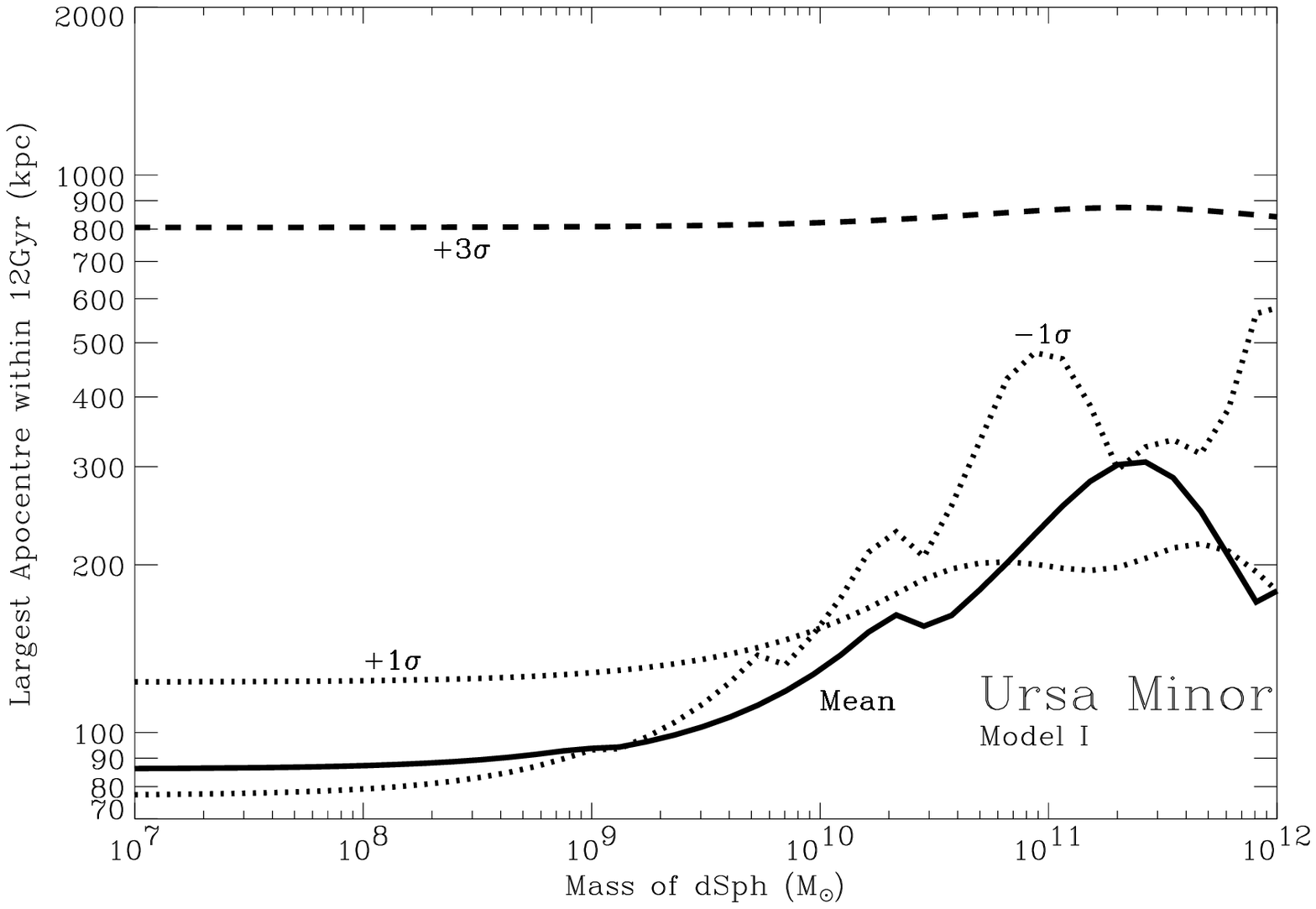}
}
\subfigure{
\includegraphics[angle=0,width=8.0cm]{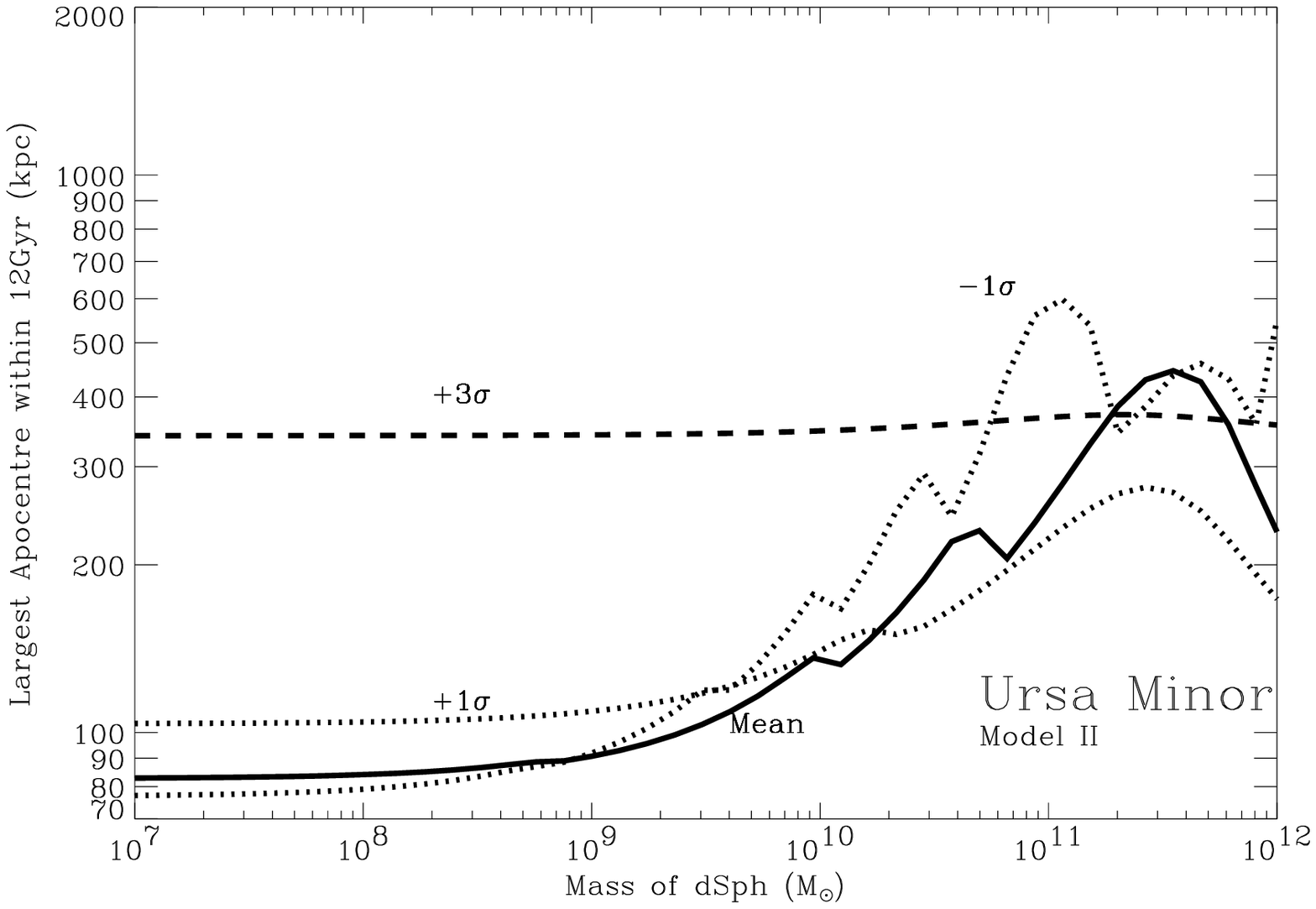}
}\\
\subfigure{
\includegraphics[angle=0,width=8.0cm]{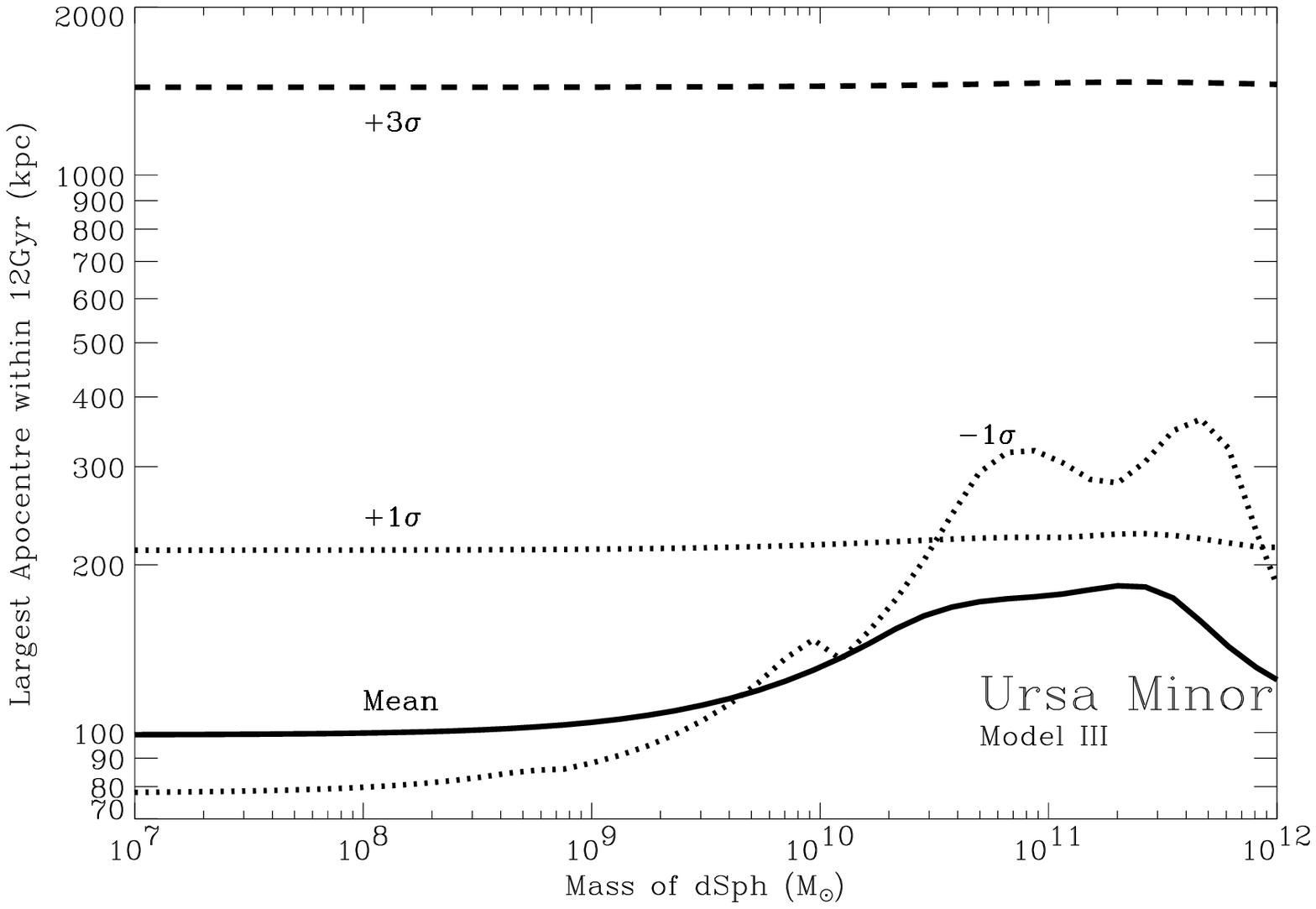}
}
\subfigure{
\includegraphics[angle=0,width=8.0cm]{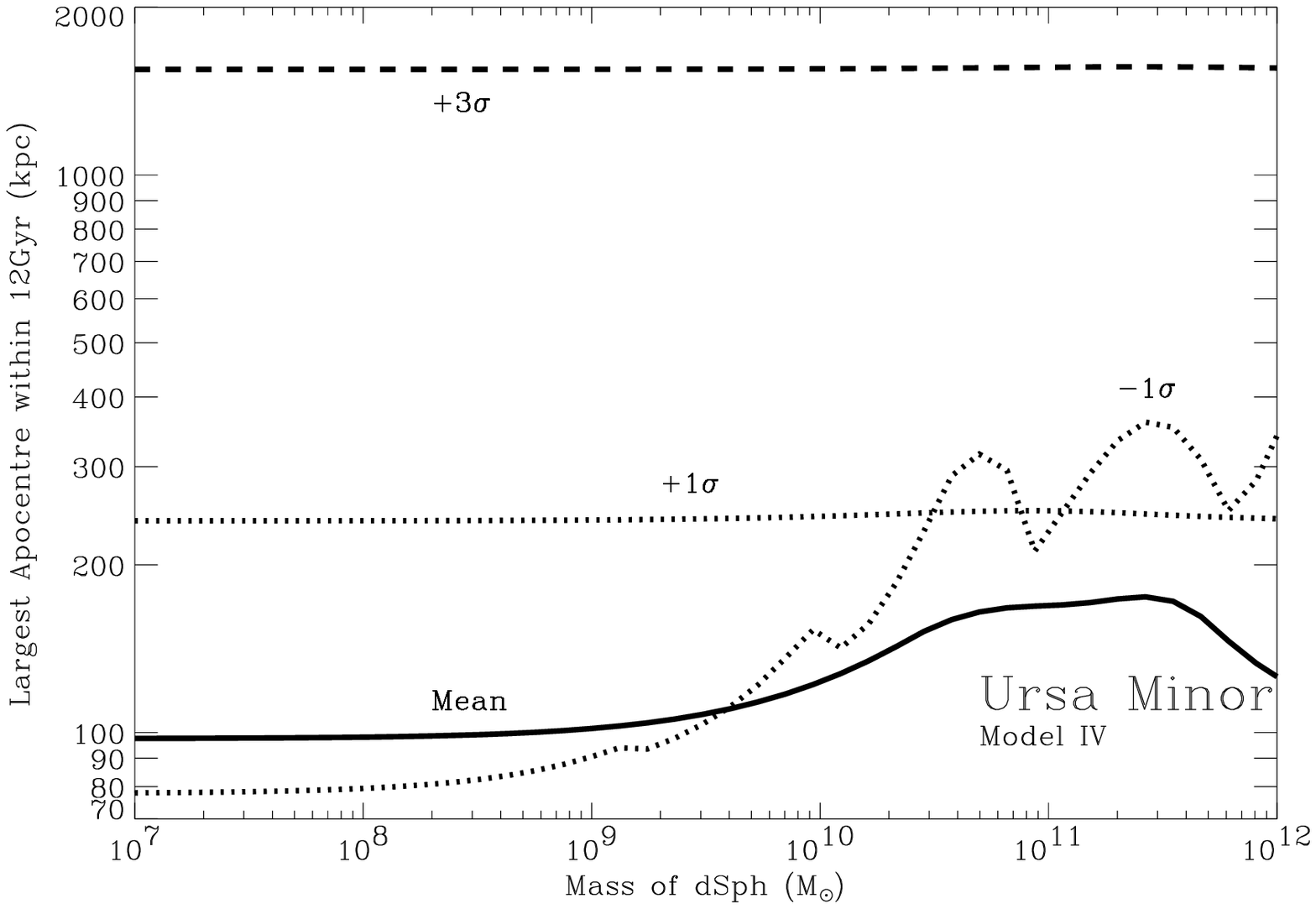}
}\\
\caption{Here we plot the largest apocentre, as a function of mass, within 12~Gyrs, for the orbit of Ursa Minor in the four Galaxy models as defined in Table~\ref{tab:mod}. The different linetypes correspond to the proper motions used: the solid line is the mean proper motion as defined in Table~\ref{tab:pm}, the dotted and dashed lines correspond to $\pm 1\sigma$ and $+3\sigma$ errors respectively. Regardless of the Galactic potential, Ursa Minor would have required a mass greater than $10^{10}\msun$, even at $1\sigma$ to have fallen in along a filament.}
\protect\label{fig:umi}
\end{figure*}

In Fig~\ref{fig:umi} we plot the largest apocentre reached within $12~Gyr$ against the dSph mass for the Ursa Minor dSph making use of all four Galactic potential models. The different linetypes, for Fig~\ref{fig:umi} - \ref{fig:hha}, correspond to the proper motions used: the solid line is the mean proper motion as defined in Table~\ref{tab:pm}, the dotted and dashed lines correspond to $\pm 1\sigma$ and $+3\sigma/-X\sigma$ errors respectively. There is a slight complication to this because in the cases of Carina and Ursa Minor, it makes no sense to consider proper motions as low as $3\sigma$ below the mean measurement. The reason for this is that using the mean measurement minus $1\sigma$ for Carina and Ursa Minor respectively, the pericentre is $\sim 10~kpc$. We stop at $10~kpc$ because as one can see from Fig~\ref{fig:orbit}, the pericentre does not vary considerably during a Hubble time, rather it is the apocentre that decays. Therefore, if a dSph had orbited to $10~kpc$, we would expect it to be tidally destroyed. 

The closer a dSph approaches, the more efficient DF is, thus the easier it is to be captured. Interestingly, if the proper motions are found to be at the higher end of the error range, it is impossible - regardless of mass - to fall in and finish with the correct position and velocities.

It is obvious from Fig~\ref{fig:umi} that there is a clear trend in all Galactic potentials for the largest apocentre to increase with dSph mass. This trend simply derives from the proportionality between the DF deceleration and the satellite mass (Eq~\ref{eqn:df}) - see also \cite{tormen98}. Regardless of the Galactic potential, Ursa Minor would have required a mass greater than $10^{10}\msun$, even at $1\sigma$ below the mean, to have fallen in along a filament from more than $200~kpc$.

\subsection{Dependence on Proper Motions}
\begin{figure*}
\centering
\subfigure{
\includegraphics[angle=0,width=8.0cm]{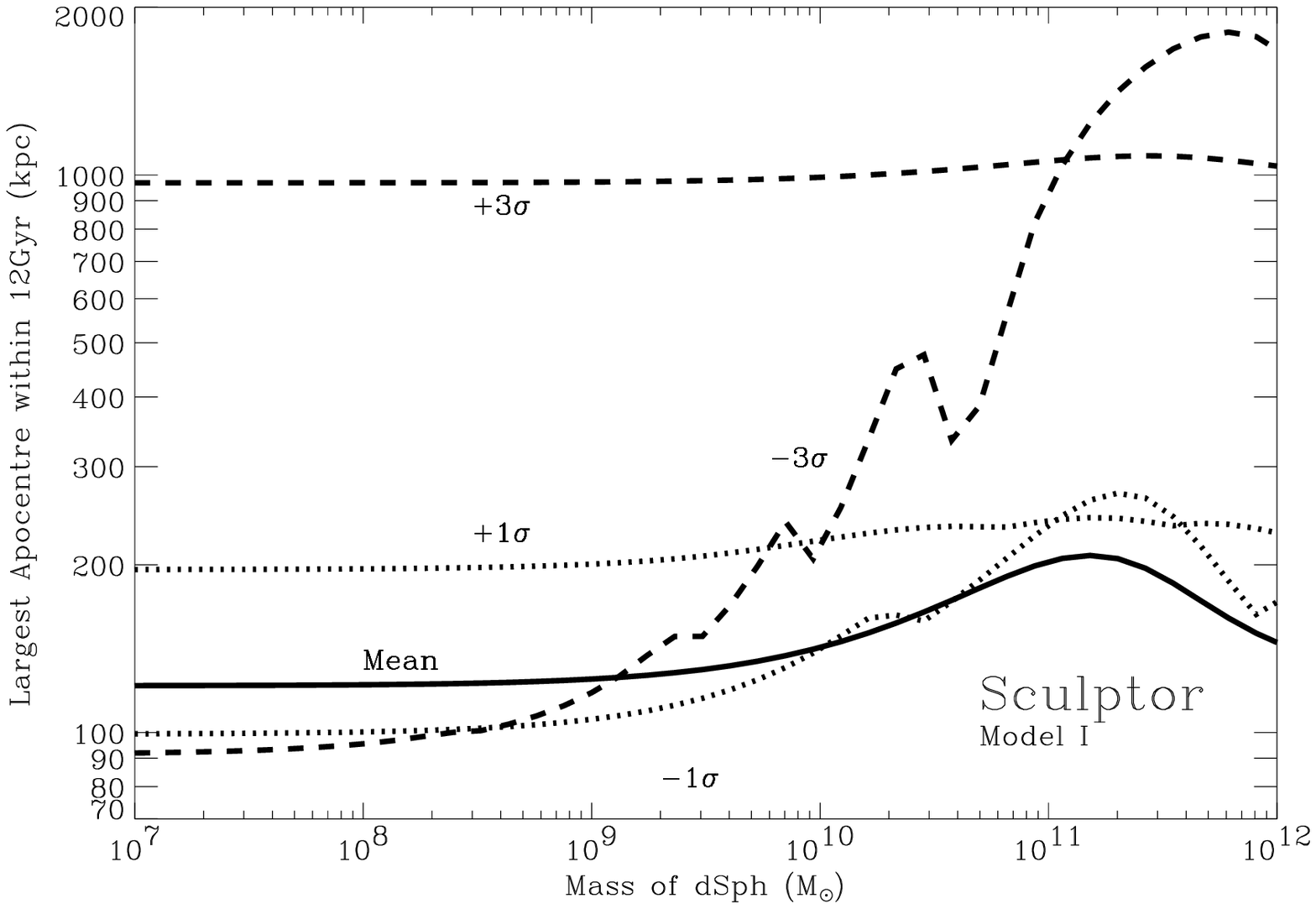}
}
\subfigure{
\includegraphics[angle=0,width=8.0cm]{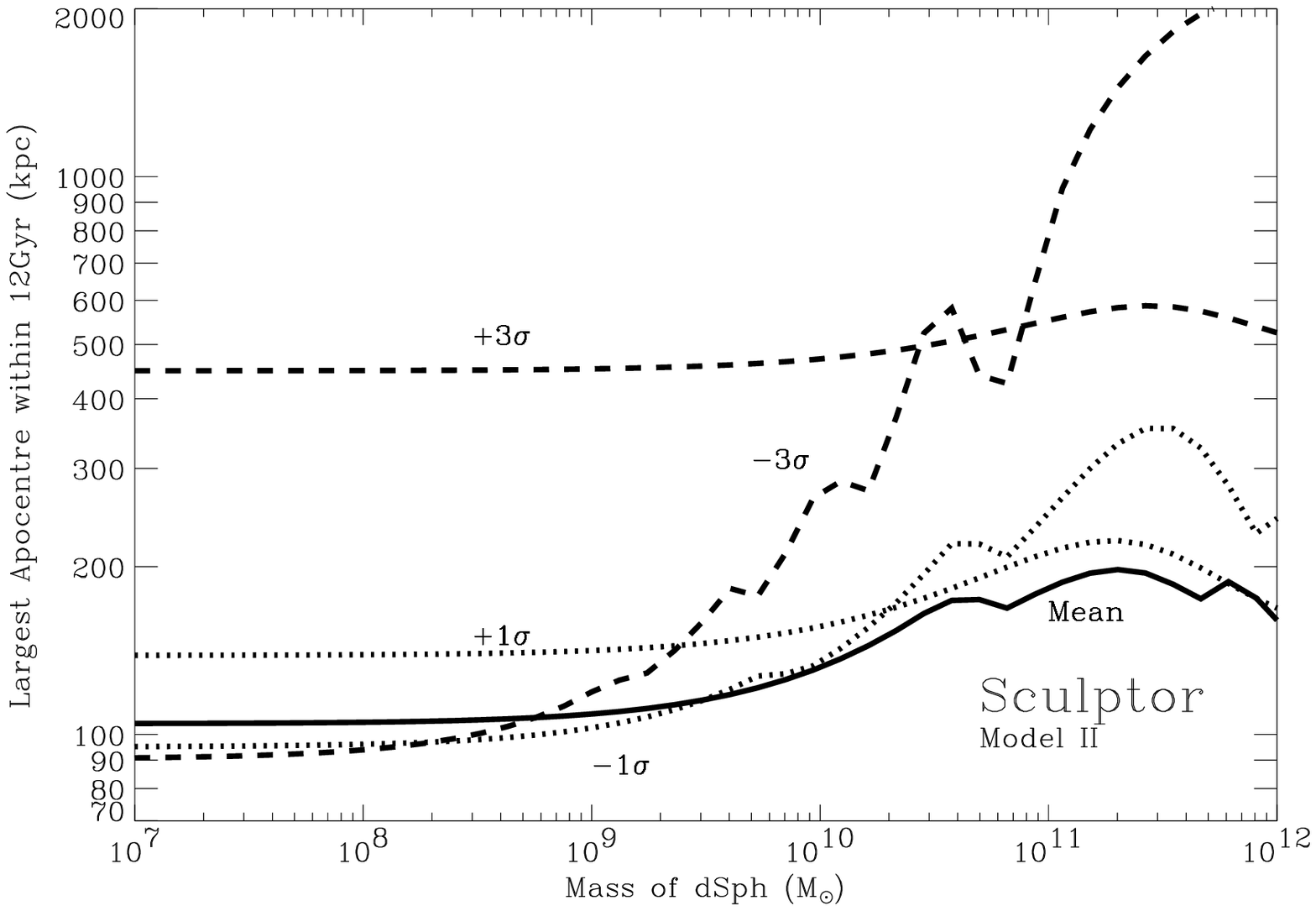}
}\\
\subfigure{
\includegraphics[angle=0,width=8.0cm]{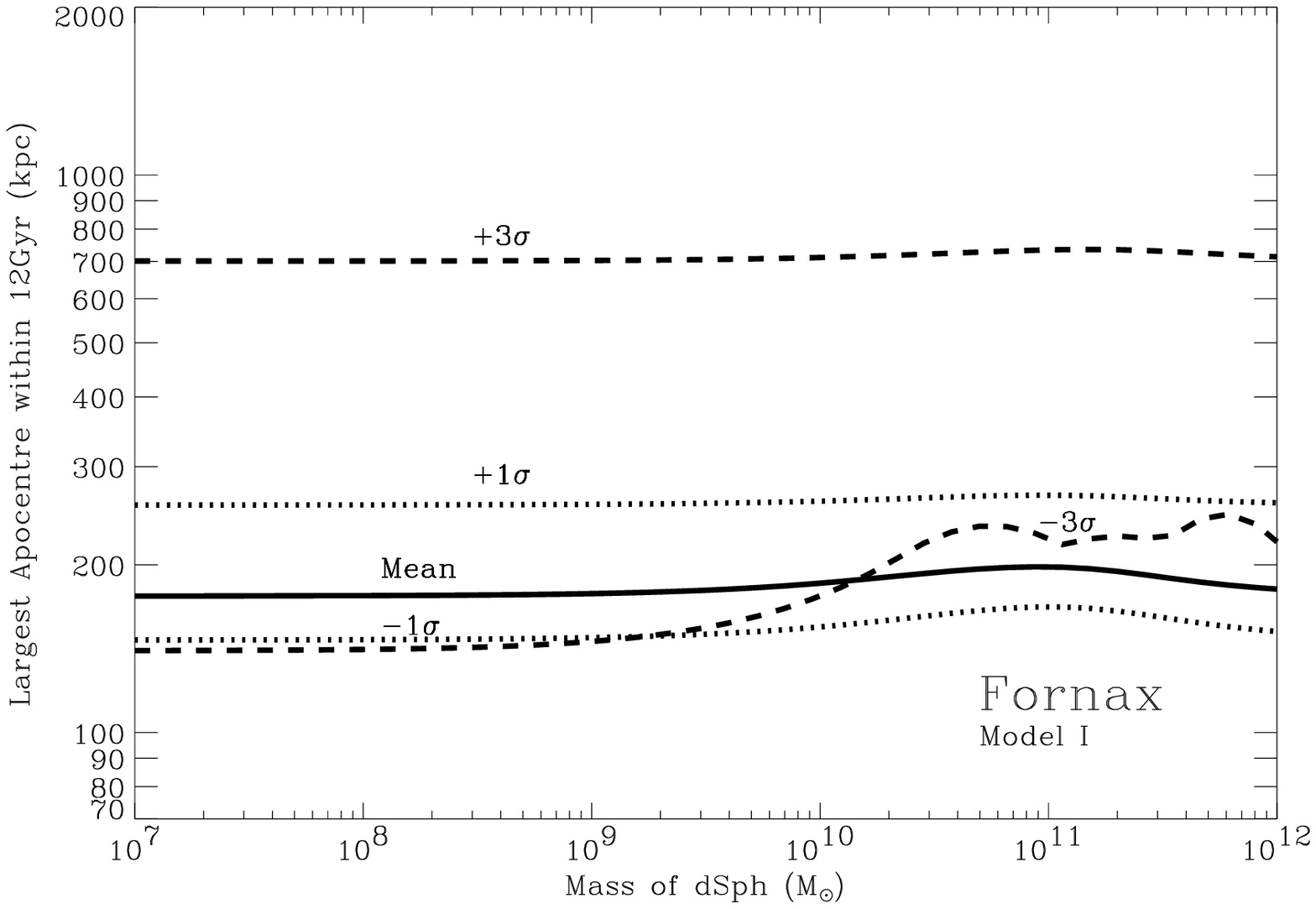}
}
\subfigure{
\includegraphics[angle=0,width=8.0cm]{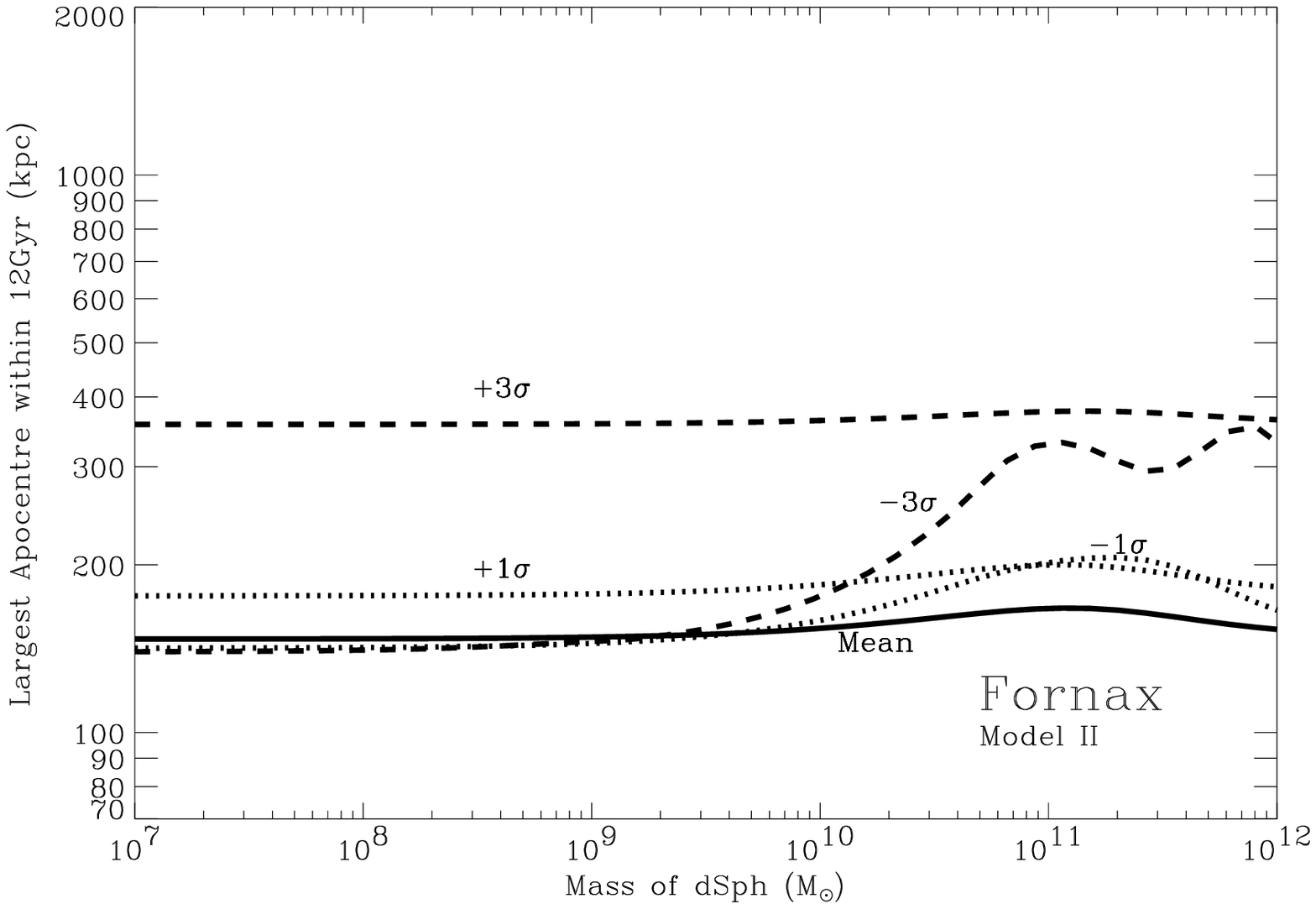}
}\\
\subfigure{
\includegraphics[angle=0,width=8.0cm]{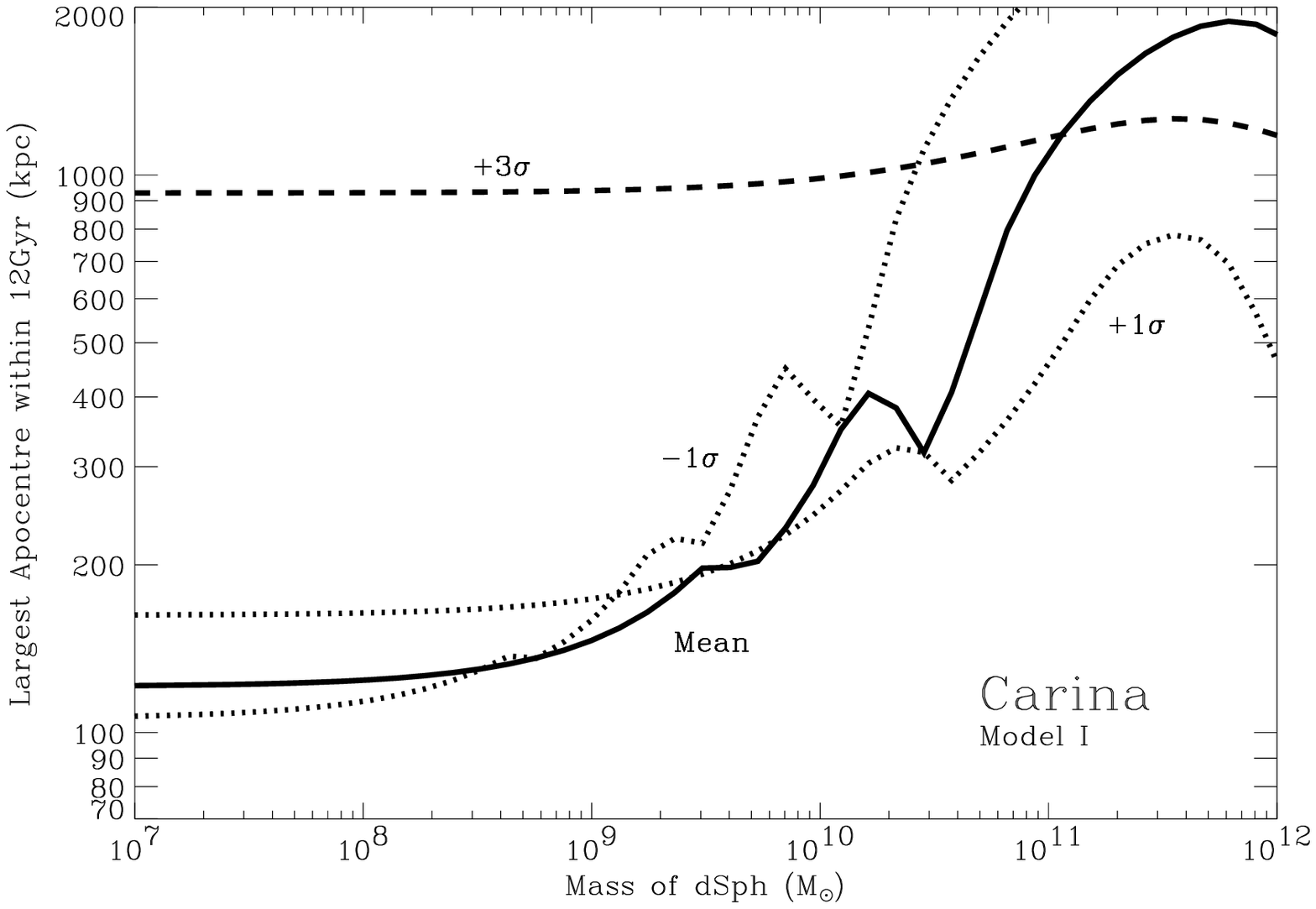}
}
\subfigure{
\includegraphics[angle=0,width=8.0cm]{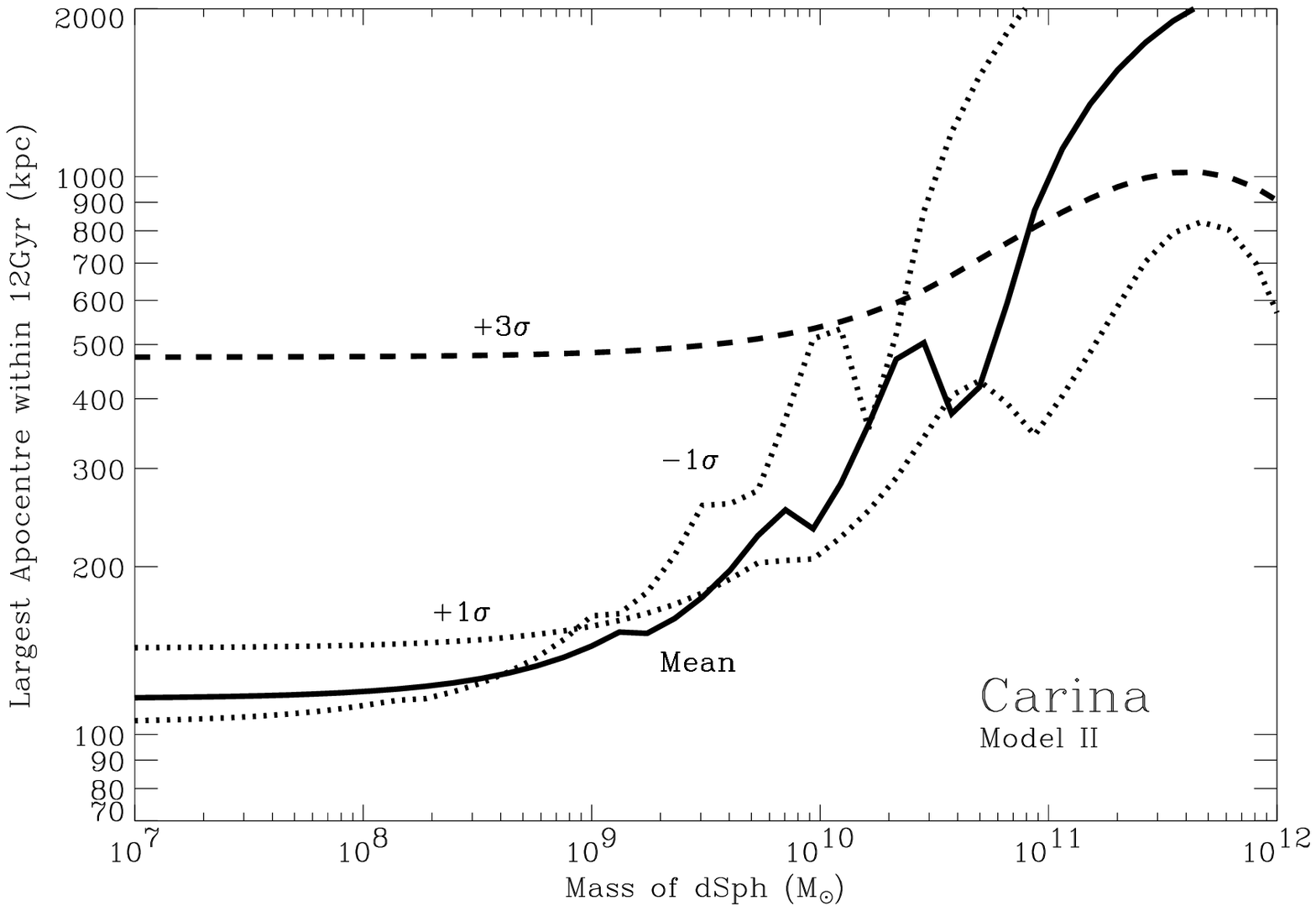}
}\\
\caption{Here we plot the largest apocentre, within 12~Gyrs, for the orbits of Sculptor, Fornax and Carina with Galaxy models I and II - as defined in Table~\ref{tab:mod}. The different linetypes correspond to the proper motions used: the solid line is the mean proper motion as defined in Table~\ref{tab:pm}, the dotted and dashed lines correspond to $\pm 1\sigma$ and $\pm3\sigma$ errors. Carina has only one dashed line as explained in \S\ref{sec:longorbs}. }
\protect\label{fig:car}
\end{figure*}

\protect\label{sec:longorbs}
Given that the Galactic model only weakly influences the ability to lose orbital energy, we plot in Fig~\ref{fig:car} the largest apocentre versus dSph mass for each of the dSphs, but only for Galactic models I and II. 

Consider Fornax and Sculptor with Galactic model I (as well as Ursa Minor with Models I and III, Fig~\ref{fig:umi}). For the $+3\sigma$ proper motions (where the proper motion error is added in parallel to the mean) the largest apocentre is constant with dSph mass. In this case, the dSph can always orbit beyond $700~kpc$. We obtain a similar result for the $+1\sigma$ where the largest apocentre is now between $200$ and $300~kpc$. The largest apocentre also is almost independent of the dSph mass for Models III and IV, which are not shown, because these models have significantly lower circular velocities beyond $30~kpc$ (cf. Fig~\ref{fig:vc}) hence lower escape speeds. 

In contrast, it is the tendency of the majority of the cases where the proper motion is subtracted from the mean measurement in Fig~\ref{fig:car} to increase their maximum apocentre with increasing dSph mass. In fact, the orbits referring to these lines begin $12~Gyr$ ago with large apocentres, but DF, whose efficiency increases with mass, reduces their orbital energy and they finish with much shortened apocentres. This is a far more plausible scenario than the orbits corresponding to the +3-$\sigma$ lines. In this latter case, the dSph would still today be orbiting to beyond $700~kpc$ and this is likely to be inconsistent with the positions of the eight classical dSphs, which are observed within $\approx 250~kpc$ of the Milky Way. In fact, out of the four we investigate here, three are within $90~kpc$, with Fornax still only at $138\pm8~kpc$. Therefore, the probability for both Fornax and Sculptor to be on a long orbit to $700~kpc$ and being observed at $138~kpc$ and $87~kpc$ is small. Thus, it is reasonable to consider only scenarios where the dSphs today have relatively small apocentres (say less than 200~kpc), hence excluding Galactic models III and IV. In addition, we can exclude Model II for Ursa Minor, Sculptor and
Fornax (top-right panel of Fig~\ref{fig:umi} and right hand panels of Fig~\ref{fig:car}), because they have never been beyond $500~kpc$ from the Milky Way 
over the last 12~Gyr. 

It would appear, therefore, that there is no scenario which can reconcile Fornax with the filamentary infall, regardless of the proper motion, assuming it is within $3\sigma$ of the mean measurement.

In the case of Sculptor, to have fallen in from far outside the virial radius, using the most favourable proper motion at $3\sigma$ below the mean, the dSph would need to weigh more than $10^{10}\msun$ for Models I and II and at $1\sigma$ it is barely possible regardless of dSph mass. 

Ursa Minor and Carina would both require masses of the order of $10^{10}\msun$ and these masses come with the condition that the proper motions are at least $1\sigma$ below the mean measurement. The reason for this, in contrast to Fornax, is that lowering the proper motions results in very low pericentres allowing much orbital energy to be shed.

A subtle point worth noting is why the largest apocentre is not a monotonically increasing function of dSph mass. This is linked to the number of pericentric orbits. By increasing the dSph mass slightly, the number of orbits can decrease by one, hence even though the DF strength at pericentre increases, the number of times it orbits to pericentre within $12~Gyr$ has dropped.

\subsection{Significance of dSph mass}
\protect\label{sec:dmass}

In the previous section we confirmed that, using this simple model, there is very little chance that Fornax fell into the Milky Way's potential along a filament. Furthermore, as long as the proper motions of Sculptor, Carina and Ursa Minor are greater than or equal to the mean, it is very doubtful that their orbits could result from filamentary infall. However, if the proper motions co-conspire to be low at more than $1\sigma$, there is the possibility that dSphs with masses of the order of $10^{10}\msun$ could have fallen into the Milky Way's potential and ended up with their current positions and velocities.

The problem with this scenario is that it is unlikely that the dark matter halos of Sculptor, Ursa Minor and Carina are so massive. In \cite{angus10}, cored dark matter halos were fitted to line-of-sight velocity dispersions of \cite{walker07}. At $2~kpc$ the enclosed masses of Carina, Fornax and Sculptor were shown to be 3, 3 and $2\times 10^{8}\msun$, Ursa Minor was not studied, but has a similar mass profile (cf. \citealt{walker07}). The cored density profiles have the same asymptotic density profile as NFWs (i.e. $\propto r^{-3}$) and hence the cumulative masses at the virial radii are not too different. In fact, the NFW fits from \cite{walker07} have masses at $2~kpc$ that are approximately 0.8, 3.2 and 2$\times 10^{8}\msun$ respectively.

Taking Carina, if we assume that it has a totally unphysical density profile such that the density stays constant at the $2~kpc$ value and integrate the total mass, we can find the minimum extent of the dark matter halo. Since the minimum dSph mass for filamentary infall is $10^{10}\msun$ and $\rho(2~kpc)=2.8\times 10^{6} \msun kpc^{-3}$, we find that $M_{d}={4\pi \over 3} \rho r^3$ meaning $r \approx 10~kpc$. This is of course the best case scenario and it is more likely that the mass would need to be integrated beyond $20~kpc$. Such diffuse halos would never survive the first pericentre passage and moreover, the DF formula (Eq~\ref{eqn:df}) would be overestimating the effect for small impact parameters.

\subsection{Host Halo Accretion}
\protect\label{sec:hha}
Given that we integrate backwards in time to $12~Gyrs$ ago, it is not satisfactory to exclude the accretion of mass to the host halo.

To be able to elucidate the effect of this on our results, we followed the model of \cite{wechsler02} where the total mass of a cold dark matter halo is seen to grow with decreasing redshift. They suggest a prescription where the mass at any given redshift, $M(z)$, is related to the mass $M_{200}$ at redshift zero by $M(z)=M_{200}e^{-\alpha z}$ to which we impose a floor on the minimum mass, such that $M(z)$ cannot drop below 5\% of $M_{200}$. Here $\alpha$ is the parameter that controls the rate of accretion and $M_{200}$ is the virial mass at redshift zero. 

If we recall that $\rho_c={3 H(z)^2 \over 8\pi G}$ then we can equate the virial radius with the accretion rate by $r_{200}^3(z)={GM_{200}e^{-\alpha z} \over 100 H(z)^2}$. Due to the accretion, the gravity imposed by the Milky Way varies in proportion to both $e^{-\alpha z}$ and a factor coming from the growing virial radius with decreasing redshift. The parameters that control the DF - $v_c^2$ and the dark matter density $\rho$ (see Eq~\ref{eqn:df}) - vary as a consequence of this prescription and we also impose the same accretion rate on the stellar mass of the Milky Way (described in \S\ref{sec:4m}).

\begin{figure*}
\centering
\subfigure{
\includegraphics[angle=0,width=5.5cm]{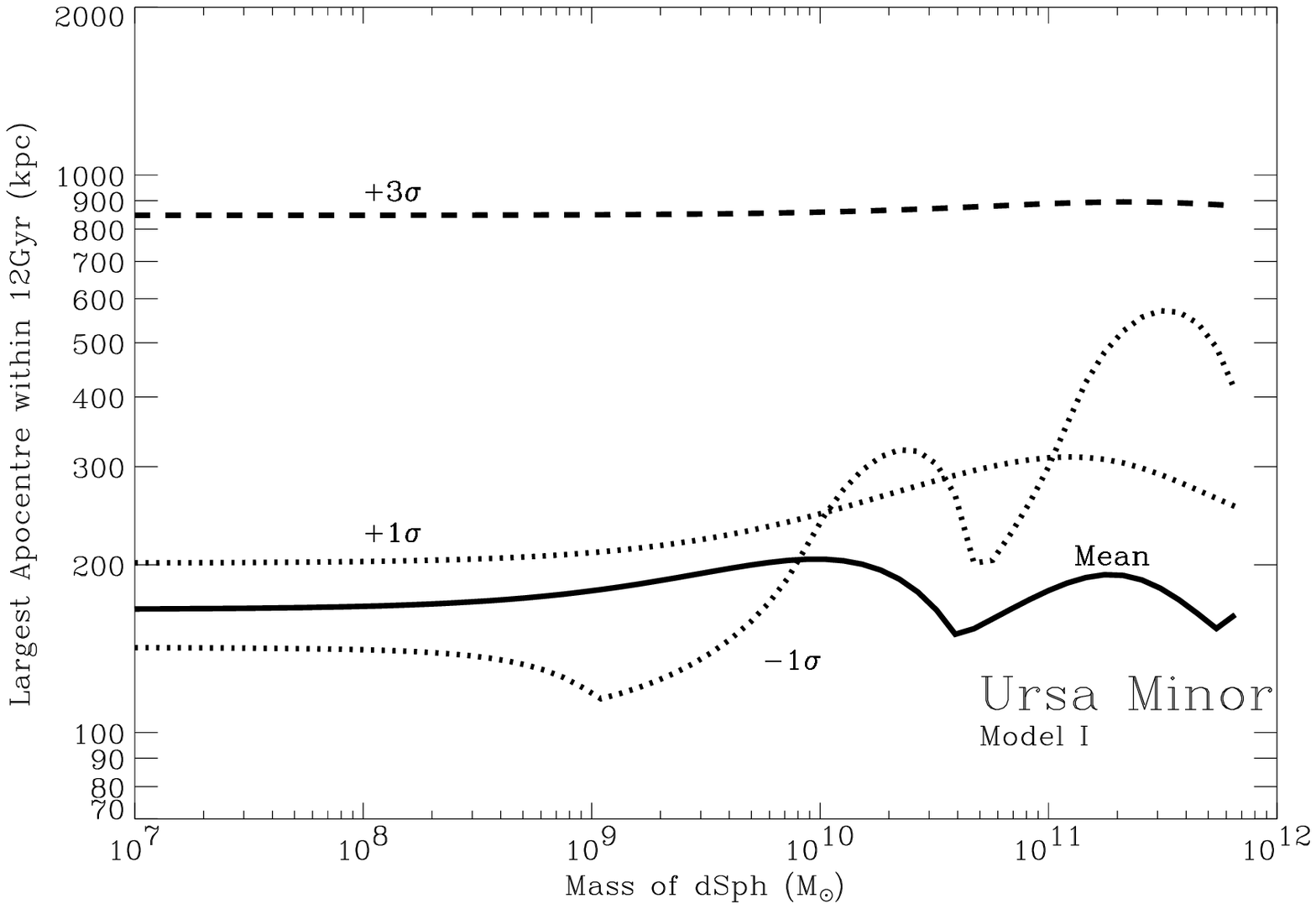}
}
\subfigure{
\includegraphics[angle=0,width=5.5cm]{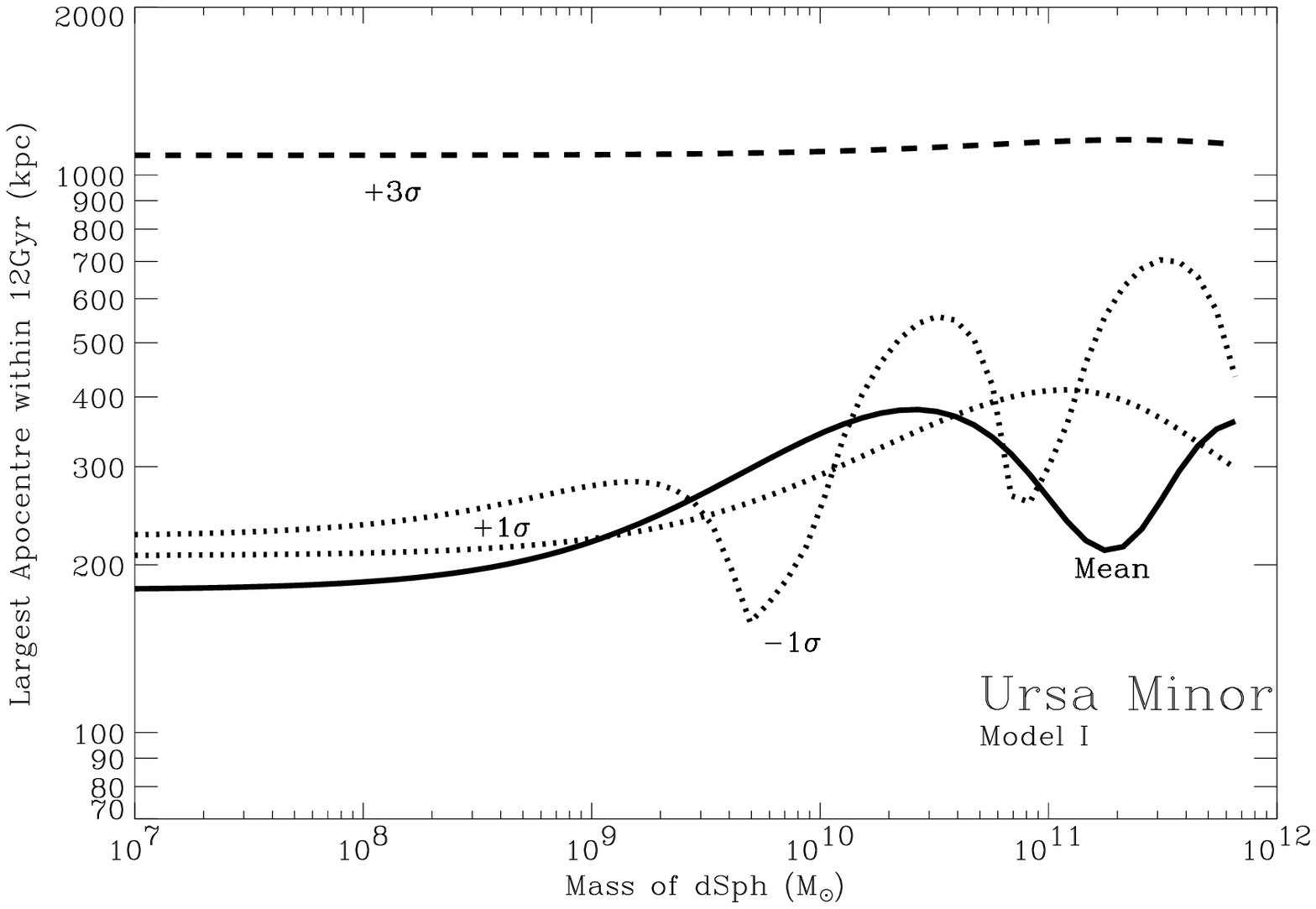}
}
\subfigure{
\includegraphics[angle=0,width=5.5cm]{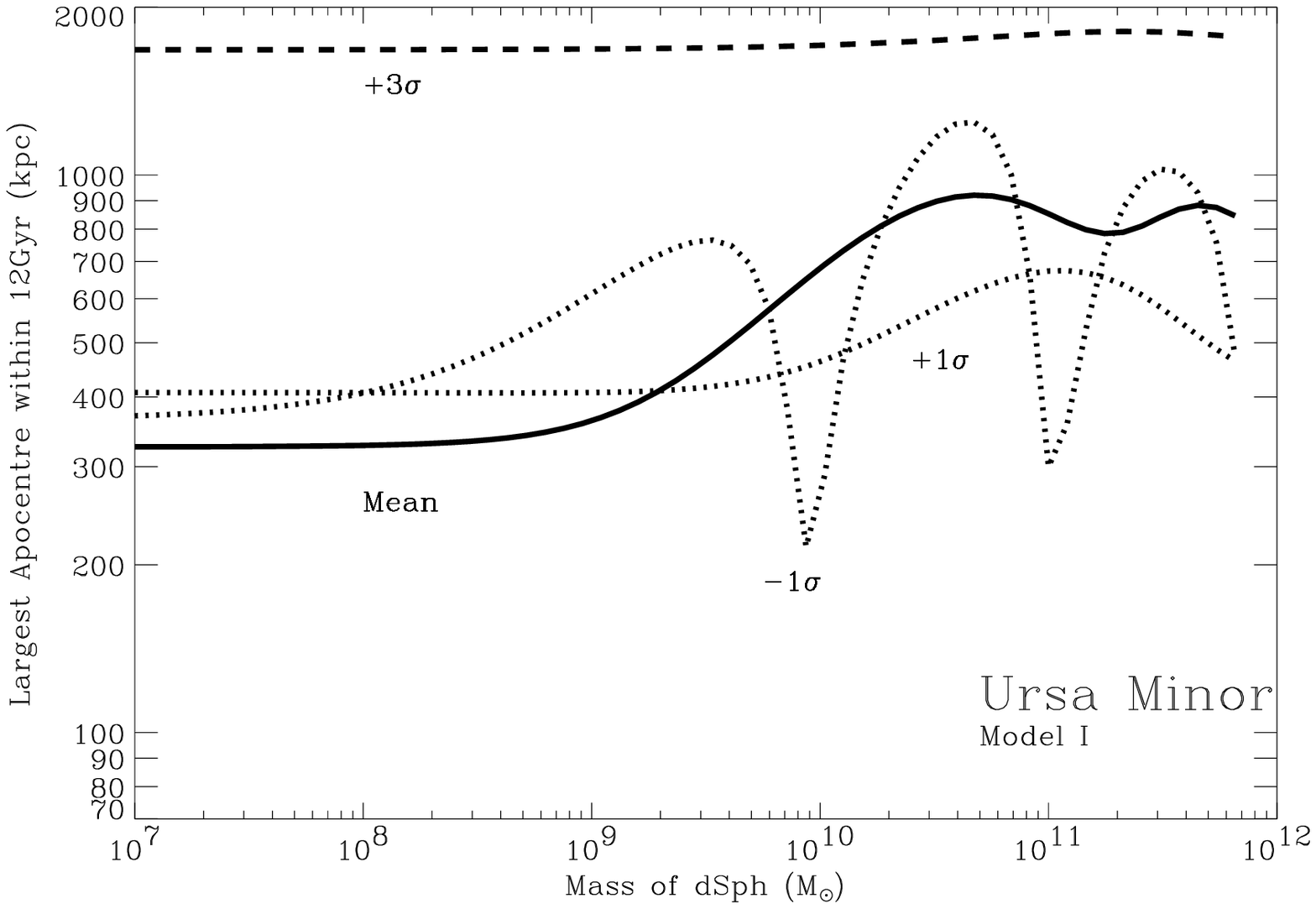}
}\\
\subfigure{
\includegraphics[angle=0,width=5.5cm]{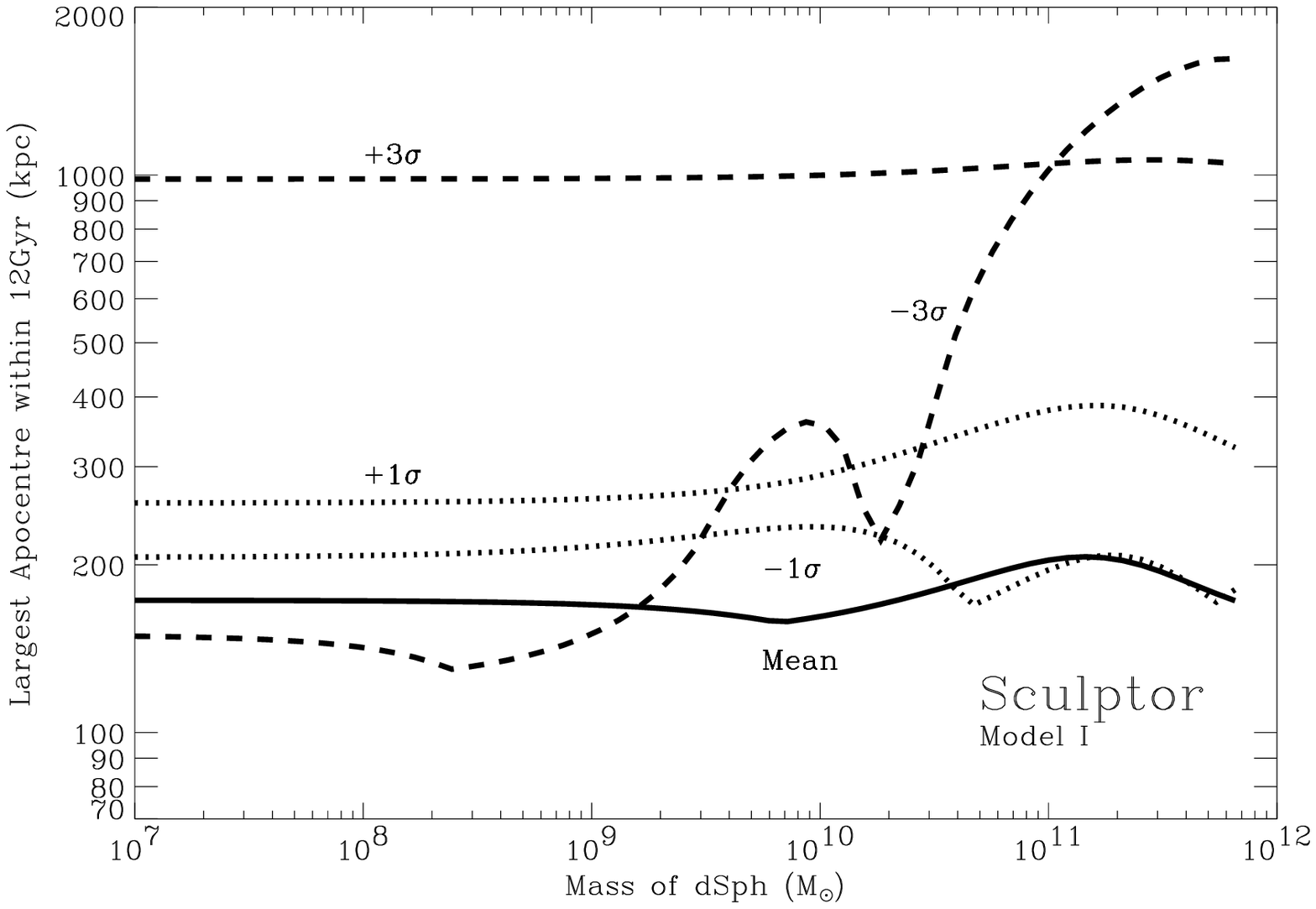}
}
\subfigure{
\includegraphics[angle=0,width=5.5cm]{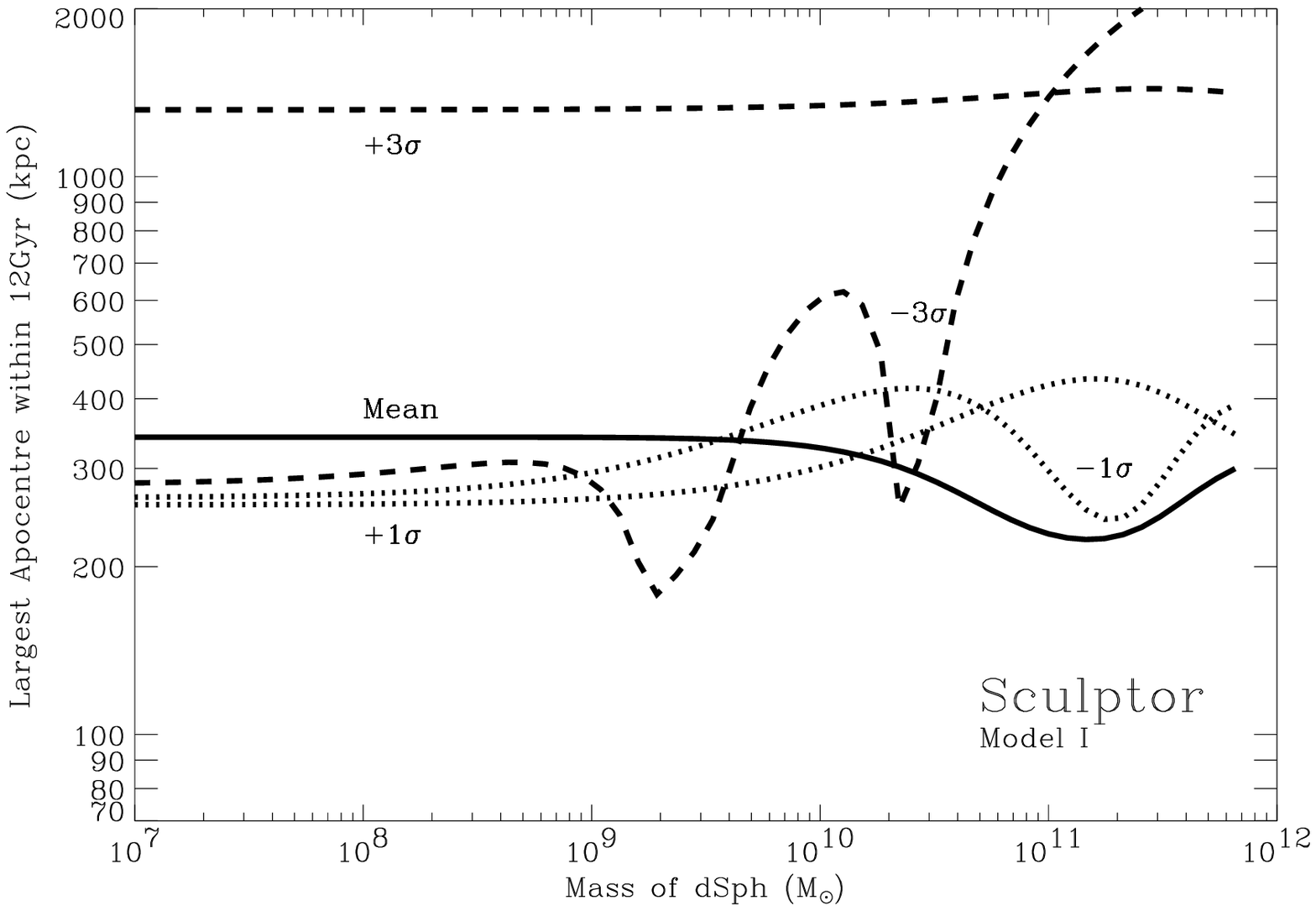}
}
\subfigure{
\includegraphics[angle=0,width=5.5cm]{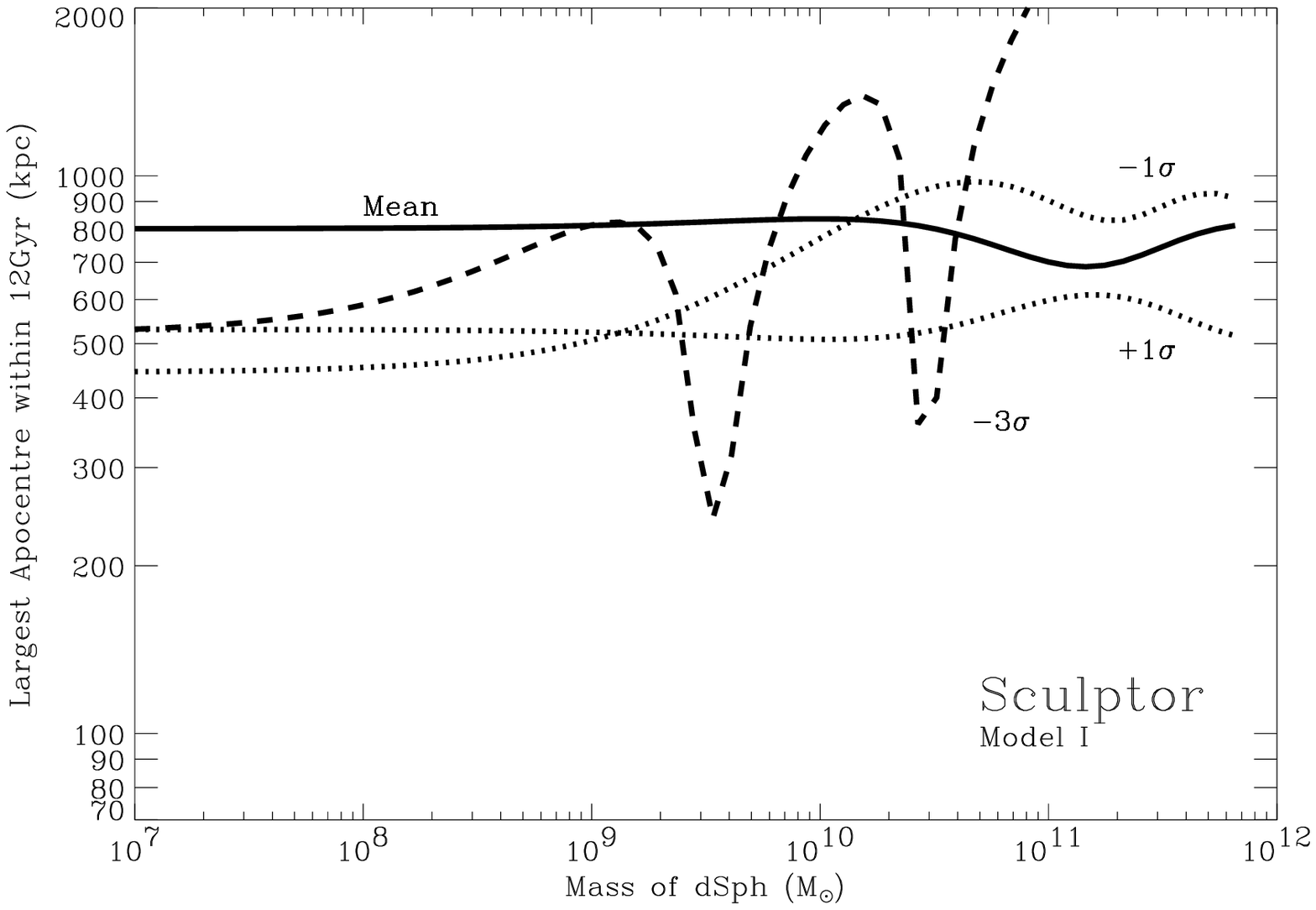}
}\\
\subfigure{
\includegraphics[angle=0,width=5.5cm]{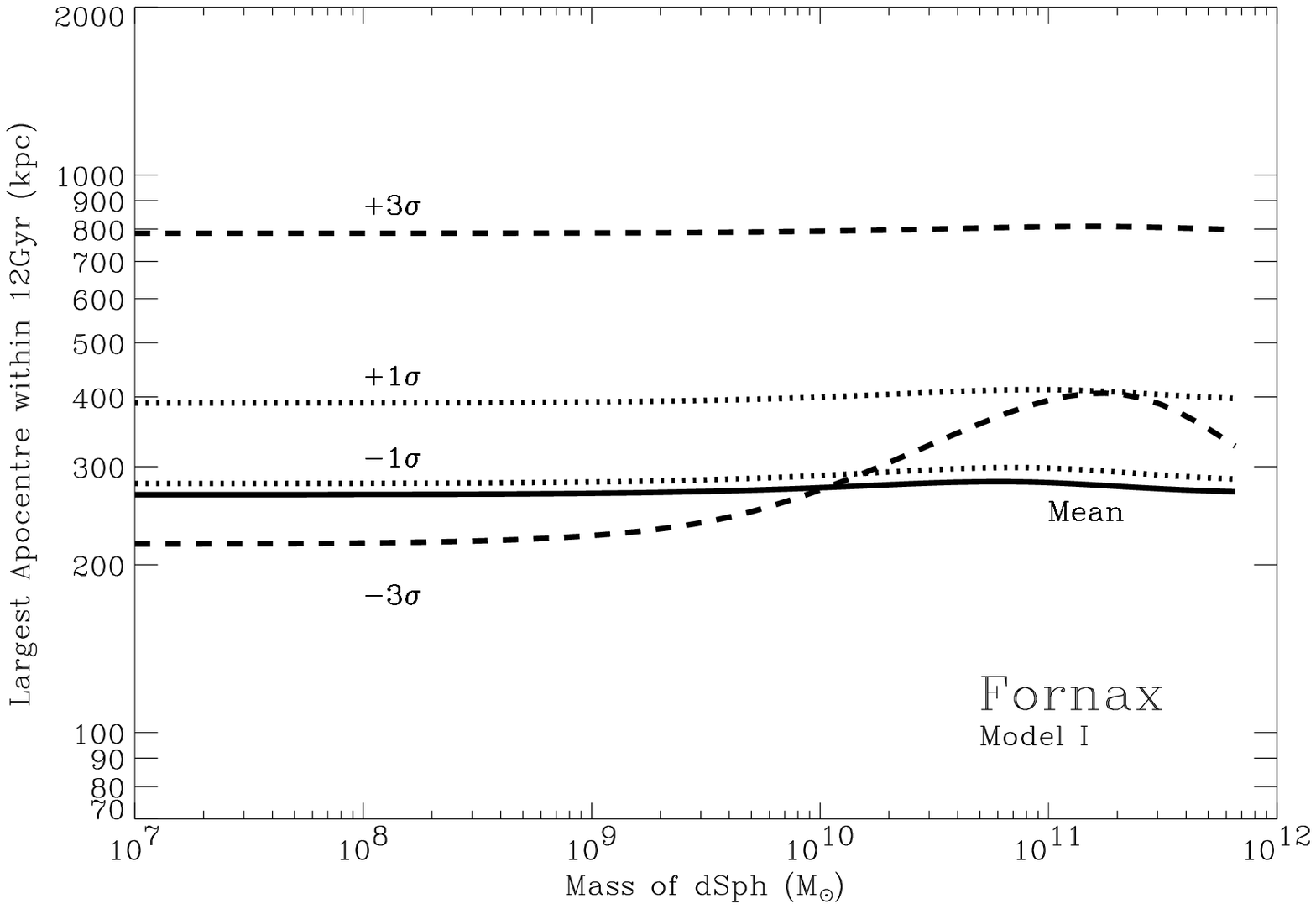}
}
\subfigure{
\includegraphics[angle=0,width=5.5cm]{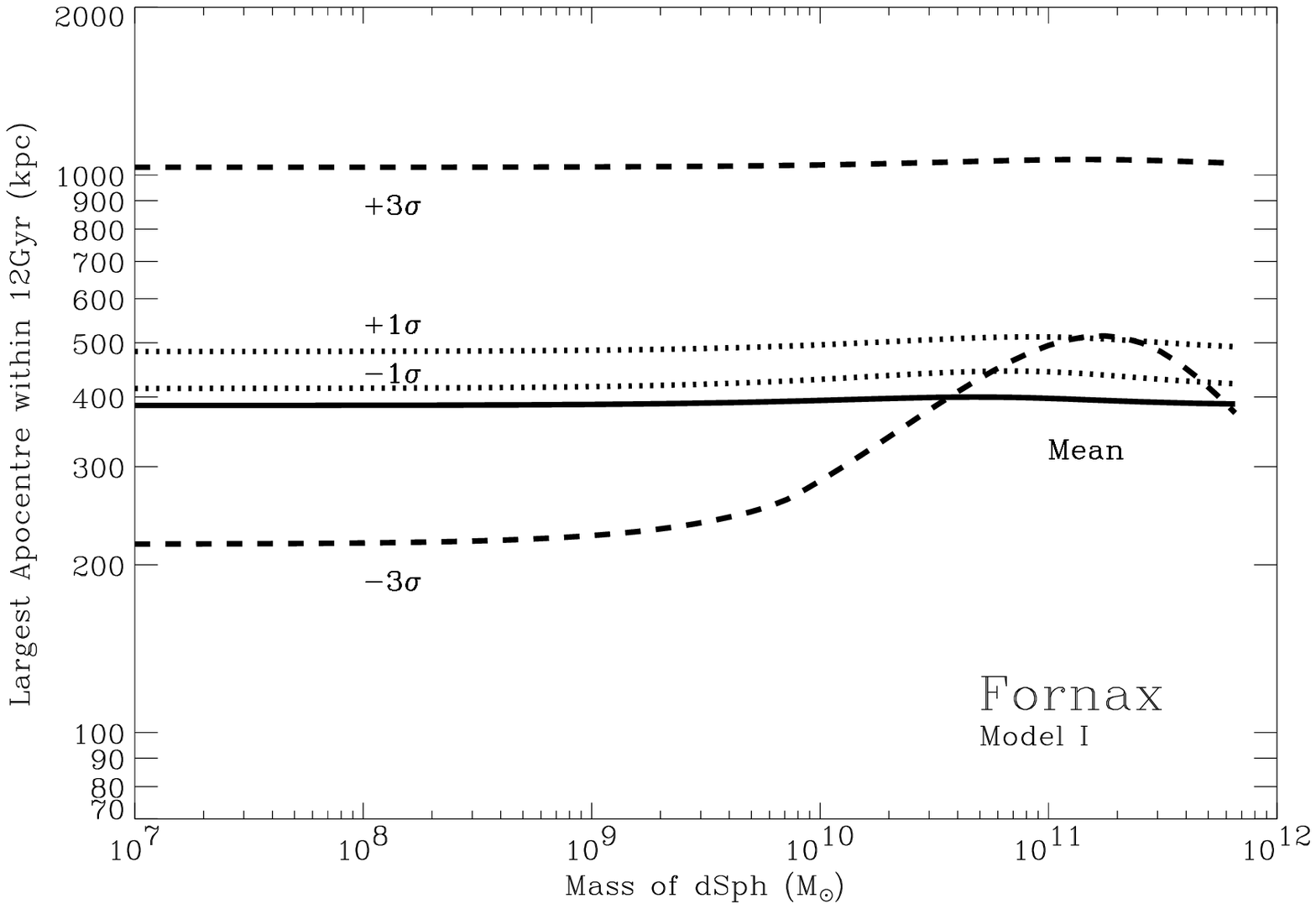}
}
\subfigure{
\includegraphics[angle=0,width=5.5cm]{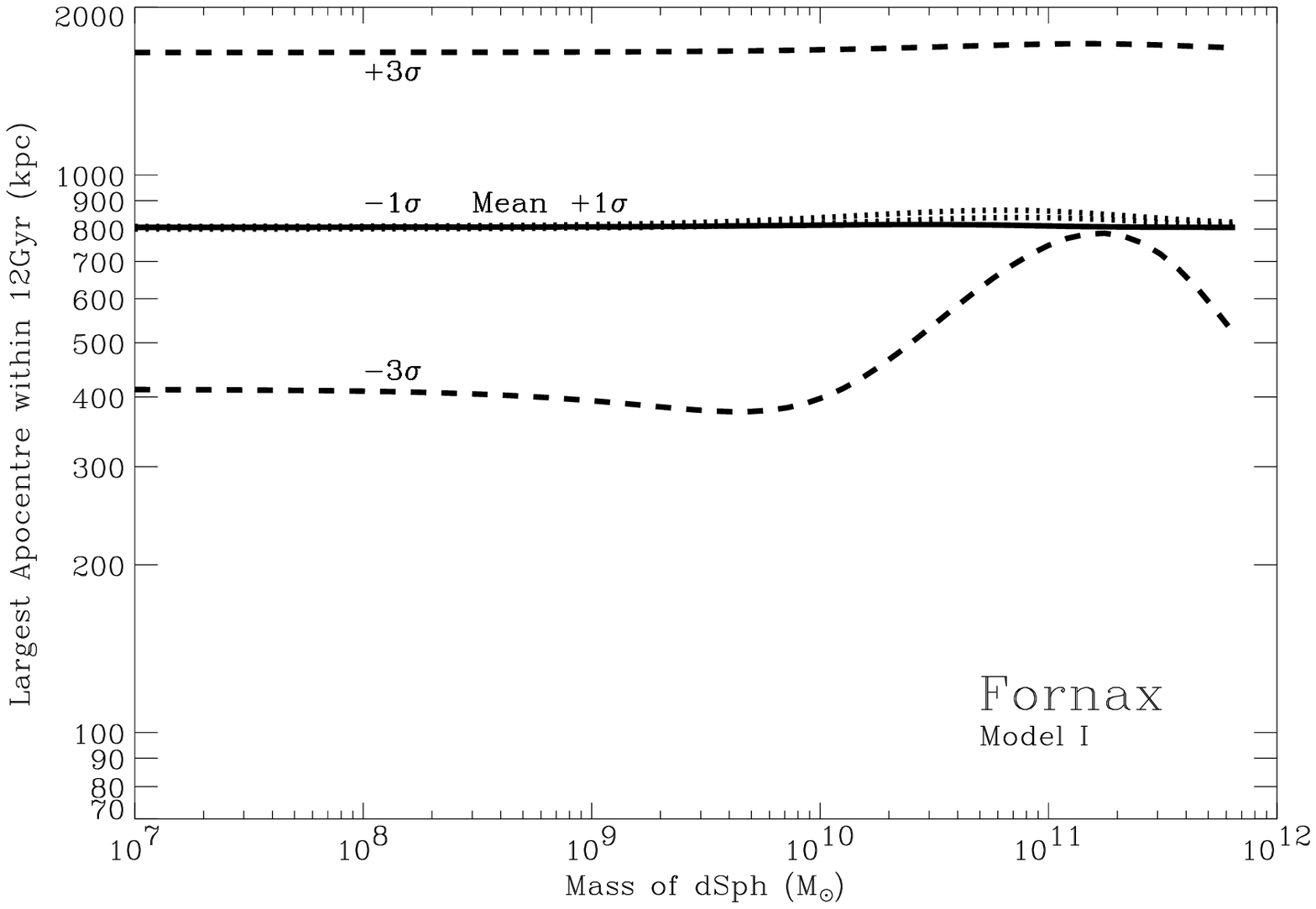}
}\\
\subfigure{
\includegraphics[angle=0,width=5.5cm]{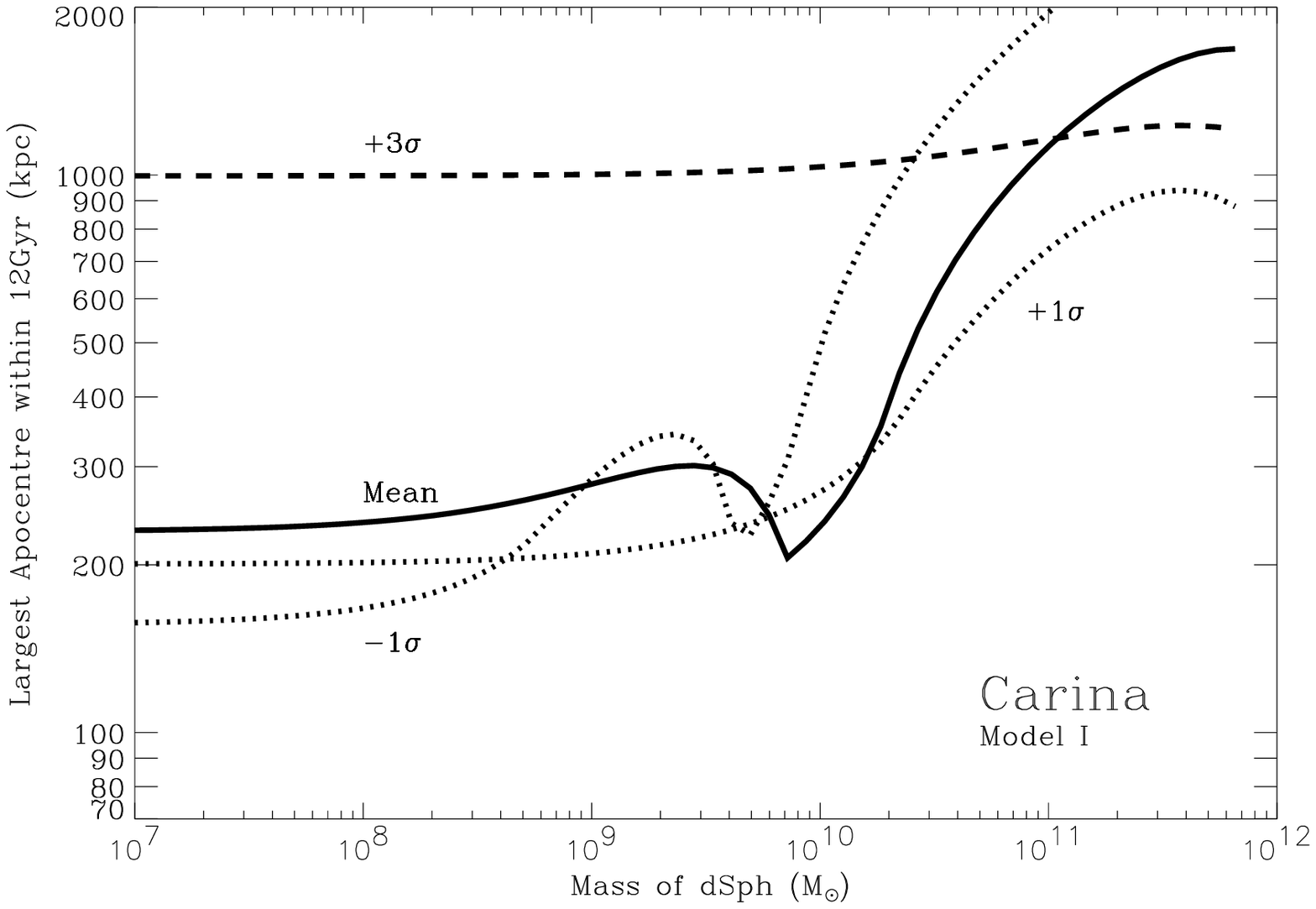}
}
\subfigure{
\includegraphics[angle=0,width=5.5cm]{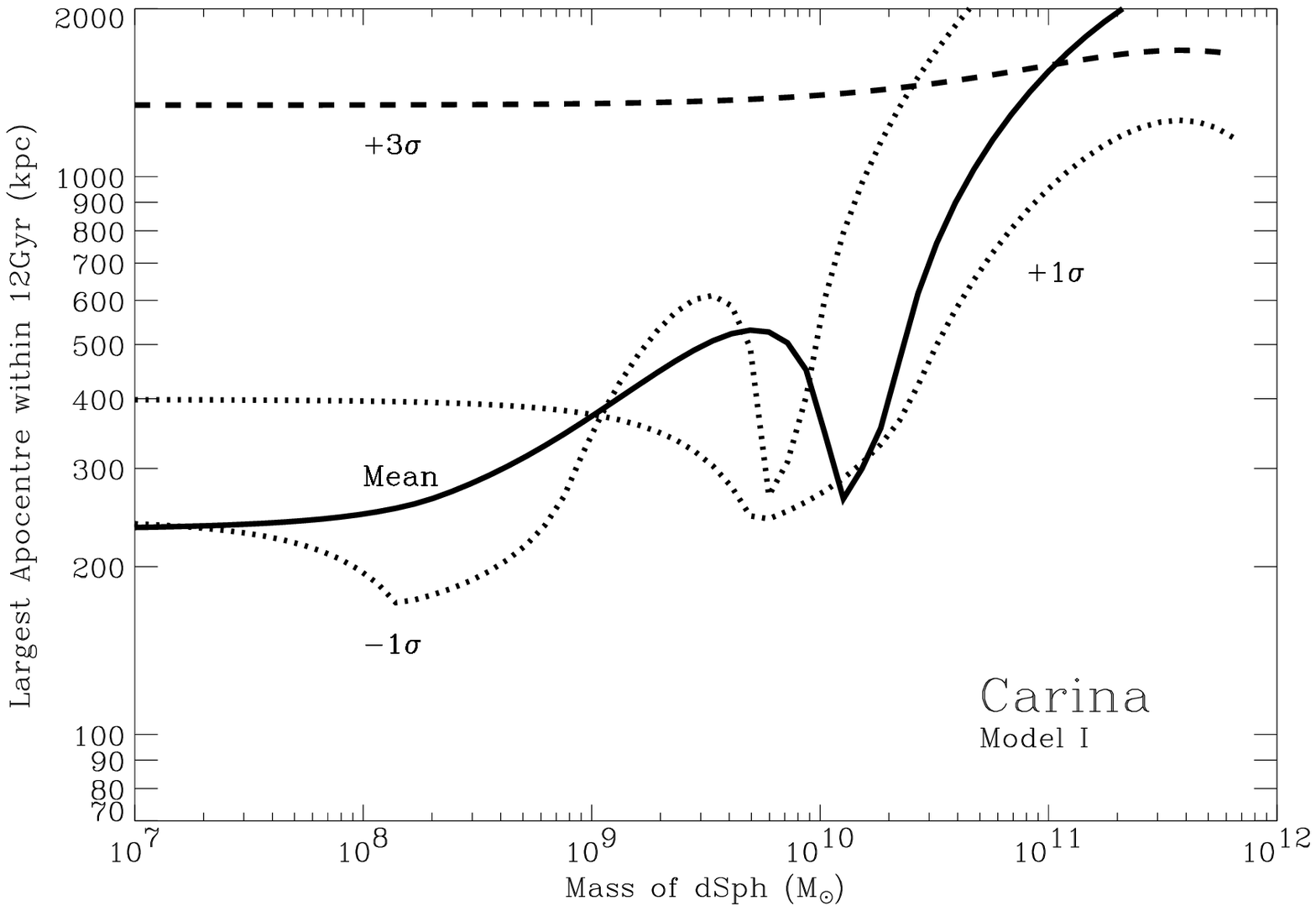}
}
\subfigure{
\includegraphics[angle=0,width=5.5cm]{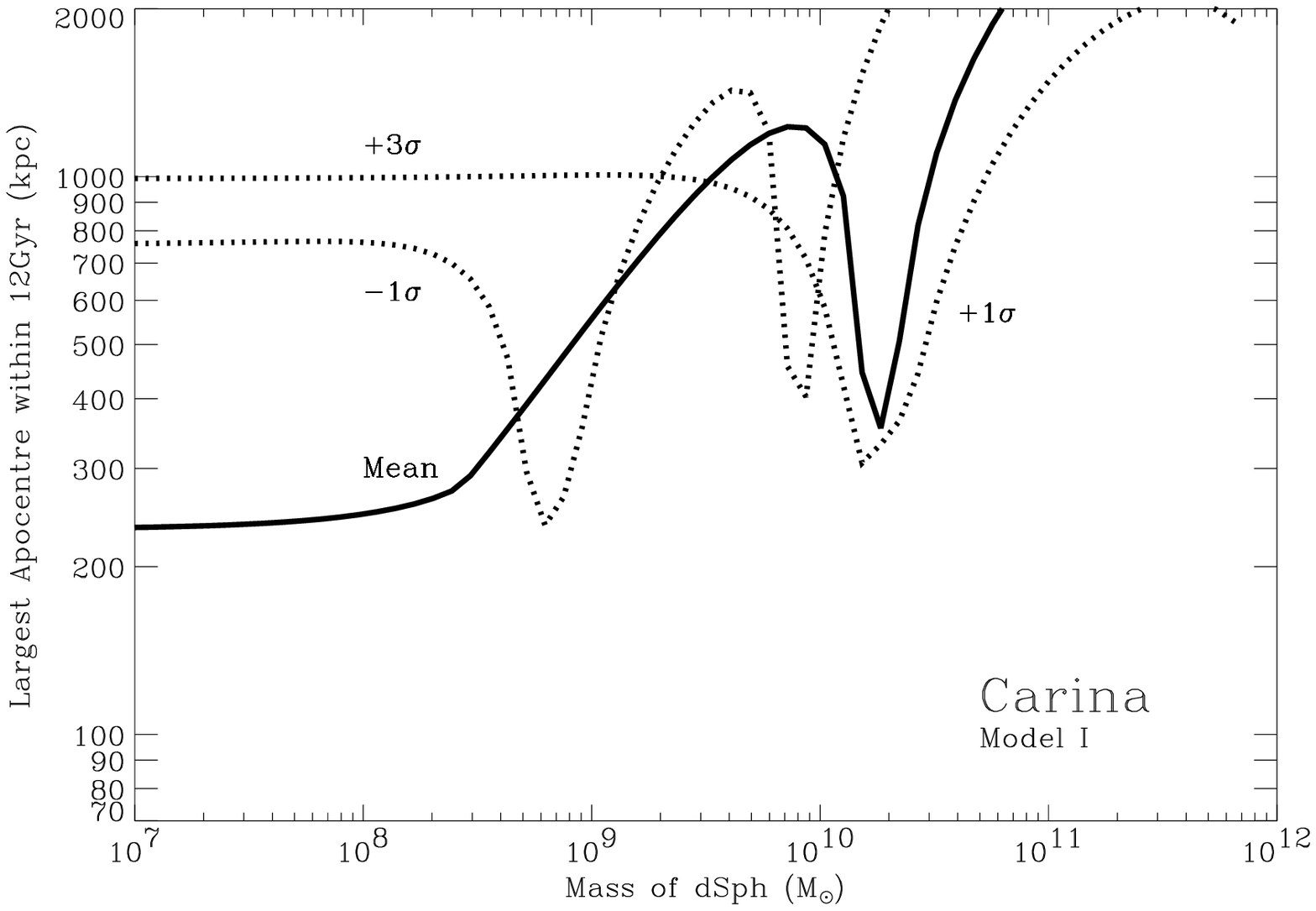}
}\\
\caption{Like for Fig~\ref{fig:car} except we include the effects of host halo accretion (see \S\ref{sec:hha}). All panels use $\alpha=0.63$, but the left, middle and right hand panels are the largest apocentre after 5, 7 and 12~Gyrs respectively. We only plot Galactic model I, however, the results are the same for model II. }
\protect\label{fig:hha}
\end{figure*}

It is suggested that typical values of $\alpha$ for Milky Way sized halos are 0.5-0.7. We tested $\alpha$ fully in the range 0 to 0.87 and plot in Fig~\ref{fig:hha} the results for $\alpha=0.63$ for Carina, Fornax, Sculptor and Ursa Minor, whilst limiting ourselves to Galactic model I. The left, middle and right hand panels show the maximum apocentre within 5, 7 and 12~Gyrs respectively.

At 5 Gyrs (see left hand panels of Fig~\ref{fig:hha}) and within 1-$\sigma$ of the mean measurement of the proper motions, the results described
in the previous sections remain valid, because there is only a slight difference between the maximum apocentres found here and those in Fig~\ref{fig:car} that were derived considering a 12~Gyrs time interval and no accretion. Most apocentres are larger than those shown in Fig~\ref{fig:car} because of the weakening potential with increasing redshift, rather than to the fact that the dSphs are escaping. Excluding Fornax, for masses smaller than $10^{10}$~M$_\odot$ the dSphs are only orbiting up to 250~kpc.

The maximum apocentre of Fornax itself, at +1-$\sigma$ from the mean measured proper motion, does increase significantly by 5~Gyrs. In Fig~\ref{fig:hha}, left hand panel, one can see the +3-$\sigma$ line of Fornax reaches 400~kpc at low masses, whereas in Fig~\ref{fig:car} the same line reaches only 250~kpc. By 7~Gyrs (middle panel of Fig~\ref{fig:hha}) this line has reached 500~kpc suggesting it could be in the throes of escaping.  In addition, at 3-$\sigma$ from the mean measurement, all four dSphs could already have escaped within 5~Gyrs, although this scenario would suffer the same flaws as stated in \S\ref{sec:longorbs} and \S\ref{sec:dmass}: either the dSphs are unlikely to be observed within $\sim 140$~kpc of the Milky Way when their orbits have apocentres
larger than $\sim 700$~kpc or the required large mass of their dark matter halo is inconsistent with current mass estimates.
{sec:longorbs}
By 7~Gyrs (the middle panels of Fig~\ref{fig:hha}) the maximum apocentres have increased somewhat. Ursa Minor's maximum apocentres only increases very mildly, however the +1-$\sigma$ line for Carina shoots up from 200kpc to 400kpc. This is a criteria for escape since its apocentre was stable at 200~kpc for 5~Gyrs and between 5 and 7~Gyrs its apocentre increased by a further 200~kpc. Other dSphs that show apocentre increases of at least 200~kpc are Sculptor (at -3-$\sigma$ and at the mean proper motion) and Carina (at +1-$\sigma$ from the mean proper motion), but certainly not Ursa Minor. The crucial point to notice here is that this only begins to happen after 5~Gyrs.

Within 12~Gyrs (right hand panels of Fig~\ref{fig:hha}) it is clear that the dSphs can reach a significant distance from the Milky Way. One can see that by 12~Gyrs for the Carina dSph, it is possible to reach 1~Mpc from the Milky Way, even with a reasonable mass ($10^8\msun$). Likewise, Fornax and Sculptor can reach 800~kpc, but Ursa Minor can only reach 500~kpc for a reasonable combination of proper motion and mass. 

Considering that 12~Gyrs ago the virial radius of the Milky Way is only a fraction of its current value, it can be concluded that all the dSphs can infall, but not more recently than 5~Gyrs ago. Whether reaching 500~kpc or 1~Mpc from the Milky Way in 12~Gyrs corresponds to a large enough separation to be compatible with filamentary infall is not ascertainable with these simulations. Alternatively, they may all have proper motions larger than the mean measurement by more than $1-\sigma$ and it could merely be a coincidence that these four dSphs are found within 140~kpc of the Milky Way whilst maintaining the long orbits to the outer halo they would have as a result of their large proper motions. The feasibility of the former scenario, namely an early infall, is not studied here, but has been extensively discussed and dismissed in \cite{klimentowski10}, because the organized distribution of the dSphs we observe today is unlikely to survive such a long period of time.


\section{Conclusion and Discussion}
Here we took the positions and velocities of the four classical dwarf spheroidal galaxies of the Milky Way that have measured proper motions, and sought to discover if these final conditions were consistent with the filamentary infall model proposed by  \cite{kang05,libeskind05,libeskind07,libeskind09,zentner05} to offer a non-convoluted explanation of the so-called disk of satellites (\citealt{kroupa05,metzkroupa07,metz09,lb83}).

Using an analytic prescription of dynamical friction, and a simple leap-frogging time step updater with negative time steps, we were able to trace back the orbit for different masses of dwarf spheroidal galaxies in various Milky Way potentials. 

It was found that the proper motions and dwarf spheroidal masses are the two key parameters and that the Milky Way potential model was rather subsidiary. The proper motions were varied up to $\pm 3\sigma$ from the mean measurement and all reasonable dwarf masses were considered.

We argued that no conditions could be found such that Fornax might be consistent with the simple filamentary infall model. Moreover, the only scenario for Sculptor, Ursa Minor and Carina to be consistent with the infall hypothesis is for their proper motions to both be significantly lower than the mean, and for them to have total masses of more than $10^{10}\msun$. The problem with this scenario is that it is in stark conflict with the measured mass profiles of the dSph using the internal line-of-sight velocity dispersions of their stellar component. Realistically, the largest mass we could expect is $10^{9}\msun$ which is not conducive to the infall scenario.

To further solidify our conclusion, we incorporated the effects of host halo accretion during the formation of the Milky Way's dark matter halo over the past 12~Gyrs. We found that this, by definition, allowed the dSphs to reach larger apocentres and allowed all dSphs to move well beyond the Milky Way's virial radius (which decreases with increasing redshift). We also showed that all four dSphs remained bound at 5~Gyrs, unless the proper motions are all incorrect at more than 1-$\sigma$. Therefore, the filamentary infall scenario strongly depends on the dSphs falling in at redshifts larger than unity when the Milky Way is in its infancy. This scenario has been studied by \cite{klimentowski10} amongst others and is not favoured due to the randomisation of the orbits of the many dSphs over the large intervening time.

There is the possibility that a cold dark matter halo weighing more than $10^{10}\msun$, that bound together several dwarf galaxies fell in, and in that way the dwarfs could lose enough orbital energy. This however has the same problem as the groups discussed in \cite{metz09} and \cite{klimentowski10}.

In addition to these points, we wish to point out certain logical discrepancies in some attempts to solve the Disk of Satellites problem. \cite{deason11} demonstrate that the anisotropic spatial distribution and angular momentum bias of the satellites at z=0 can be attributed to them being accreted relatively recently compared to the majority of the dark matter of the Milky Way. Whereas, on the other hand, \cite{nichols11} claim to be able to account for the radial distribution of the dSphs if and only if they are accreted between $z=3$ and $z=10$. This example demonstrates that only small aspects of the whole problem can be solved. For example that the satellite galaxies of the Milky Way are gas poor can be understood if their gas was stripped, but this then requires them to have fallen in a long time ago. On the other hand, to explain the Disk of Satellites the dSphs must have fallen in recently.

It would appear, therefore, that on the balance of this evidence, the filamentary infall model proposed to orchestrate the positions of the classical Milky Way satellites is untenable. We suggest that proper motions for the other four classical dwarf spheroidal galaxies would greatly strengthen our conclusion, as would greater accuracy on the current proper motions. Any future proposition to explain the disk of satellites in the concordance cosmological model must take care to adhere, not only to the highly organised positions of the satellites, but also to their proper motions.

It is to be emphasised though that even in the presence of new proper motion measurements, it is not clear how the current conclusions can be changed. The new proper motion measurements would need to significantly deviate from the current best values, and the masses of the satellites would have to be made at least ten times greater than observed in order to give an infall solution. Both together do not appear to be feasible, and so the conclusions reached here that the satellites cannot have fallen in appear to be robust.

A further hurdle to the infall model is that while the gas may be distributed in a thin filament, the dark matter substructures, and notably those as heavy as $10^{10}\msun$ have filament thicknesses comparable to the diameters of the dark halo hosts they are feeding. Infall of dark matter sub-structures hosting satellites is therefore more likely to be nearly isotropic, or at best will lead to a tri-axial structure but with dimensions much thicker than the $10-30~kpc$ rms {\it thickness} of the disk of satellites.

When this evidence is added together with the elegant work of \cite{peebles10} that showed the strongly contradictory numbers of galaxies in the local void and the continuing failure of cusps in low surface brightness galaxies (\citealt{deblok10}), as well as the baryonic Tully-Fisher relation for gas rich galaxies described in \cite{mcgaugh11} the deck is seriously stacked against the concordance cosmological model at the galactic scale.

In certain modified gravity theories such as Modified Newtonian Dynamics (\citealt{milgrom95} and more recently \citealt{angus08,serra10,kosowsky10}), where cold dark matter is replaced by boosting gravity in regions of weak accelerations, the possibility of a simple origin of the dwarf spheroidals can hold because they can (and must) be tidal dwarf galaxies (\citealt{kroupa97,kroupa10,bournaud07,gentile07a,milgrom07b,tiret08a}) for which the correlated positions and low orbital velocities are perfectly natural.

\begin{table}
\begin{tabular}{l|cc}
Model& $\rho_s$& c \\
& $10^6\msun kpc^{-3}$ & \\
\hline
I &  8 & 13.5 \\
II & 8 & 12.0 \\
III& 8 & 16.5 \\
IV&340& 75.0 \\&
\end{tabular}
\caption{In the left hand part of the table we present the scale densities and concentration parameters for the four Galactic dark halos. $r_200$ is always set to 275~$kpc$ at redshift zero.}
\protect\label{tab:mod}
\end{table}

\begin{table}
\begin{tabular}{l | r@{$\pm$}l r@{$\pm$}lr@{$\pm$}l} 
dSph 
	& \multicolumn{2}{c}{$r_0$ [$kpc$]} 
	& \multicolumn{2}{c}{$V_{x_0}$ ($\kms$) }
	&  \multicolumn{2}{c}{$V_{y_0}$ ($\kms$)}\\ 
\hline
Fornax      & 138 & 8 & -31.8 &1.7 & 196 & 29 \\ 
Sculptor     & 87  &4 & 79      &6     & 198&50 \\ 
Ursa Minor & 76  &4 & -75     &44   & 144&50\\ 
Carina       & 101 &5 & 113     &52    & 46&54\\ 
\end{tabular}
\caption{Here we give the Galactocentric distances and velocities of the dwarf spheroidal galaxies. For Fornax, Sculptor and Ursa Minor, our $V_{x_0}$ corresponds to Piatek et al's $V_r$ and our $V_{y_0}$ to their $V_t$. For Carina the proper motion comes directly from \citet{pasetto11}. Distances come from \citet{mateo98}.}
\protect\label{tab:pm}
\end{table}

{\bf Acknowledgements} We sincerely thank the referee for his/her careful reading of the manuscript and for suggesting the additional 
test with the mass accretion of the host halo. GWA and AD 's research is supported by the University of Torino, Regione Piemonte by Grant and the INFN grant PD51 and the PRIN-MIUR-2008 grant 'Matter-antimatter asymmetry, dark matter and dark energy in the LHC Era'.

\begin{appendix}
\section{Integration scheme}
In order to be clear about the way the dynamical friction affects the equations of motion, we write them in full here.
\bey
\nonumber
x_{i+1/2}&=&x_i+V_{x_{i}}\Delta t/2\\
\nonumber
y_{i+1/2}&=&y_i+V_{y_{i}}\Delta t/2\\
V_{x_{i+1}}&=&V_{x_{i}}-\left({x_i \over r_i}{v_c(r_i)^2 \over r_i} +{V_{x_i}\over |V_i|}a_{df}(r_i,V_{i})\right)\Delta t\\
\nonumber
V_{y_{i+1}}&=&V_{y_{i}}-\left({y_i \over r_i}{v_c(r_i)^2 \over r_i} +{V_{y_i}\over |V_i|}a_{df}(r_i,V_{i})\right)\Delta t\\
\nonumber
x_{i+1}&=&x_{i+1/2}+V_{x_{i+1}}\Delta t/2\\
\nonumber
y_{i+1}&=&y_{i+1/2}+V_{y_{i+1}}\Delta t/2\\
\eey
where $r_i=\sqrt{x_i^2+y_i^2 }$, $V_i=\sqrt{V_{x_i}^2+V_{y_i}^2 }$, and $\Delta t$ is the time step which is negative.

Note that initially we assume all dSphs are at $x=r_0$ and $y=0$, where $r_0$ is the current Galactocentric distance and therefore that $V_{x_0}$ is the current (and initial) radial component of velocity and $V_{y_0}$ is the tangential component, as given in Table~\ref{tab:pm}. The calculations are done in the plane of the orbit just to minimise the degrees of freedom, and does not affect the results in any way.
\end{appendix}

\end{document}